\documentclass[a4paper,UKenglish,cleveref, autoref, thm-restate]{lipics-v2021}

\title{A Nearly Optimal Deterministic Algorithm for Online Transportation Problem (Full Version)} 

\author{Tsubasa Harada\footnote{Corresponding author}}{Institute of Science Tokyo, Japan}{harada.t.30af@m.isct.ac.jp}{https://orcid.org/0000-0001-8101-4153}{}

\author{Toshiya Itoh}{Institute of Science Tokyo, Japan}{titoh@comp.isct.ac.jp}{https://orcid.org/0000-0002-3706-2163}{}

\authorrunning{T. Harada and T. Itoh} 

\Copyright{Tsubasa Harada and Toshiya Itoh} 



\category{} 

\relatedversion{} 

\acknowledgements{}

\nolinenumbers 

\EventEditors{Keren Censor-Hillel, Fabrizio Grandoni, Joel Ouaknine, and Gabriele Puppis}
\EventNoEds{4}
\EventLongTitle{52nd International Colloquium on Automata, Languages, and Programming (ICALP 2025)}
\EventShortTitle{ICALP 2025}
\EventAcronym{ICALP}
\EventYear{2025}
\EventDate{July 8--11, 2025}
\EventLocation{Aarhus, Denmark}
\EventLogo{}
\SeriesVolume{334}
\ArticleNo{105}

\usepackage{lmodern}
\usepackage{ascmac}
\usepackage{amsthm}
\usepackage[ruled,vlined]{algorithm2e}
\usepackage{amsmath,amssymb}
\usepackage{mathtools}
\usepackage{type1cm}
\usepackage{comment}
\usepackage{here}
\usepackage{float}
\usepackage{accents}
\usepackage{mathrsfs}
\usepackage{algorithmic}
\usepackage[enable] {easy-todo}

\newcommand{\alg}{\mathcal{A}}
\newcommand{\algs}{\mathcal{A}^*}
\newcommand{\opt}{\mathrm{Opt}}
\newcommand{\hyb}{\mathcal{H}}
\newcommand{\ofa}{\mathrm{OTR}}
\newcommand{\ommn}{\mathrm{OMM}}
\newcommand{\omm}{\mathrm{OMM}_S}
\newcommand{\ommt}{\mathrm{OMT}_S^2}
\newcommand{\start}{t_{\sf d}}
\newcommand{\tend}{t_{\sf c}}

\newcommand{\tct}{\to\cdots\to}
\newcommand{\lca}{\mathrm{lca}}
\newcommand{\parent}{\mathrm{par}}
\newcommand{\cavity}{\mathrm{Cav}}

\newcommand{\stsvr}{a_{\sf d}}
\newcommand{{\stime}}{decoupling time}
\newcommand{\etime}{coupling time}
\newcommand{{\textdecsvr}}{decoupling server}

\newcommand{\svrinst}[3]{s_{#1}(#2, #3)}
\newcommand{\svrmpfs}[3]{s^{#3}(#1, #2)}
\newcommand{\priority}[3]{\mathrm{pri}^{#1}(#2;#3)}
\newcommand{\wE}{E_w}
\newcommand{\mc}[1]{\mathcal{#1}}
\newcommand{\mf}[1]{\mathfrak{#1}}

\newcommand{\half}{\vskip.5 \baselineskip}

\begin{document}

\maketitle

\begin{abstract}
For the online transportation problem with $m$ server sites, it has long been known that the competitive ratio of any deterministic algorithm is at least $2m-1$. Kalyanasundaram and Pruhs conjectured in 1998 that a deterministic $(2m-1)$-competitive algorithm exists for this problem, a conjecture that has remained open for over two decades.

In this paper, we propose a new deterministic algorithm for the online transportation problem and show that it achieves a competitive ratio of at most $8m-5$. This is the first $O(m)$-competitive deterministic algorithm, coming close to the lower bound of $2m-1$ within a constant factor.
\end{abstract}

\section{Introduction}
\label{sec-intro}

\subsection{Background}

The \textit{online transportation problem} ($\ofa$),
also known as the \textit{online facility assignment},
was introduced by Kalyanasundaram and Pruhs~\cite{KalP1995}.
In this problem,
$k$ servers are placed at $m$ $(\leq k)$ sites on a metric space and
an online algorithm receives (at most) $k$ requests one-by-one in an online manner.
The number of servers at one site is considered its capacity.
The task of an online algorithm 
is to assign each request irrevocably and immediately
to one of the available servers.
The cost of assigning a request to a server is
determined by the distance between them.
The objective of the problem is to
minimize the sum of the costs of assigning all requests.
We denote the problem as $\ofa(k,m)$ when there are $k$ servers and $m$ server sites.

The online transportation problem finds application in various scenarios.
Here are two examples:
In the first example, we regard a server site as a hospital,
a hospital's capacity as the number of beds it has,
and a request as a patient.
This problem can then be viewed as a problem of finding a way to assign patients to hospitals so that patients are transported to the hospital as close as possible.
In the second example, we regard a server site as a car station,
a capacity of the car station as the number of cars it can accommodate,
and a request as a user of this car sharing service.
The problem can then be viewed as a problem of designing a car sharing service that allows users to use car stations as close as possible.

The \textit{online metric matching} ($\ommn$),
or \textit{online weighted matching},
is a special case of $\ofa$
in which $k$ servers are placed at distinct $k$ sites, i.e.,
each server has unit capacity.
Let $\ommn(k)$ denote $\ommn$ with $k$ servers. 
Kalyanasundaram and Pruhs~\cite{KalP1993} and Khuller et al.~\cite{KMV1994} independently
showed that for $\ommn(k)$,
the competitive ratio of any deterministic algorithm is at least $2k-1$.
This immediately leads to a lower bound of $2m-1$ on the competitive ratio of any deterministic algorithm for $\ofa(k,m)$.
Furthermore, they also proposed a $(2k-1)$-competitive algorithm for $\ommn(k)$ called \textit{Permutation} in~\cite{KalP1993}.
From this result,
the Permutation algorithm could be expected to have a matching upper bound of $2m-1$ on the competitive ratio for $\ofa(k,m)$. 
However, Kalyanasundaram and Pruhs~\cite{KalP1998network} reported without proofs that the competitive ratio of Permutation is $\Theta(k)$
and that the competitive ratio of the natural greedy algorithm is $2^m-1$.
These results imply the existence of a large gap between the upper bound $O(\min\{k, 2^m\})$ and lower bound $\Omega(m)$ on the competitive ratio for $\ofa(k,m)$.
Since $m \leq k$, when $k$ is sufficiently larger than $m$, e.g., $k=O(2^m)$, this gap becomes even more pronounced.
Based on this discussion, they posed the following conjecture.
\begin{conjecture}[Kalyanasundaram and Pruhs~\cite{KalP1998network}]
\label{conj-2m-1}
    For $\ofa(k,m)$,
    what is the optimal competitive ratio in terms of $m$?
    It seems that there should be a $(2m-1)$-competitive algorithm.
\end{conjecture}

Since the publication of this conjecture, there have been many studies of $\ommn$ and $\ofa$. Among them, Nayyar and Raghvendra~\cite{nayyar2017input} proved that the competitive ratio of the \textit{Robust-Matching} algorithm~\cite{R2016} is $O(m\log^2k)$, thereby reducing
the upper bound on the competitive ratio for $\ofa$ to $O(\min\{m\log^2k, k, 2^m\})$.
However, neither the upper nor the lower bounds on the competitive ratio have been improved since then.

\subsection{Our contributions}
\label{subsec-contribution}

In this paper, we propose a new deterministic algorithm called \textit{Subtree-Decomposition} and show that it is $(8m-5)$-competitive for $\ofa$ with $m$ server sites. This is the first deterministic algorithm achieving $O(m)$-competitiveness within a constant factor of Conjecture~\ref{conj-2m-1}, significantly reducing the upper bound on the competitive ratio for $\ofa$ from $O(\min\{m\log^2k,k, 2^m\})$ to $8m-5$. Given that the lower bound for $\ofa$ is known to be $2m-1$, our algorithm achieves the best competitive ratio in terms of its order with respect to $m$ (and $k$). Furthermore, our algorithm processes each request in $O(m)$ time, whereas the \textit{Robust-Matching} algorithm, which has a competitive ratio of $O(m\log^2 k)$, requires $O(m^2)$ time per request~\cite[Theorem 6]{R2016}. In other words, when $k$ is close to $m$ (e.g. $k=O(m)$), although the upper bounds $8m-5$ and $O(m\log^2 m)$ differ by only a poly-logarithmic factor, our algorithm is computationally more efficient than \textit{Robust-Matching}.

We also develop a generic method to convert an algorithm designed for tree metrics into one for general metric spaces without significantly degrading the competitive ratio. This method can be applied to designing an online algorithm for optimization problems on a metric space, such as $\ommn$.

\subsection{Our techniques}
\label{subsec-technique}

\subsubsection*{Reduction to a special class of \boldmath{$\ofa$} instances}

First, we will explain how to reduce the task of designing an $O(m)$-competitive algorithm for general instances of $\ofa$ to the task of designing an algorithm for more specific instances. To this end, let us begin by defining two special cases of $\ommn$ and introducing the concept of \textit{T-strong competitive ratio}, which serves as a stricter performance measure for algorithms compared to the standard competitive ratio.

The first special case of $\ommn$ is $\ommn$ where each request is placed on the same position as a server (denoted by $\omm$) and the second one is a further special case of $\omm$ called \textit{online matching on a power-of-two tree metric} (denoted by $\ommt$). In this problem, a metric space is induced by a weighted tree $T=(V(T),E(T))$ in which the weight of each edge is a non-negative integer power of two, and the set of server sites coincides with the set $V(T)$ of vertices. Let $\omm(k)$ denote $\omm$ with $k$ servers and $\ommt(T)$ denote $\ommt$ with a metric space $T$.
Note that $|V(T)|=|E(T)|+1$ is the number of servers in $\ommt(T)$.

The T-strong competitive ratio is defined for $\ofa$ on a tree metric (the `T' is derived from `tree'). While the standard competitive ratio measures both the cost of the algorithm and the cost of the optimal offline algorithm in terms of the sum of the edge weights along the path between requests and servers (commonly referred to as the `path distance'), the T-strong competitive ratio measures the algorithm's cost using the path distance, whereas the cost of the optimal offline algorithm is measured using the weight of the heaviest edge on the path between the requests and servers (referred to as the `max-weight distance'). In other words, we say that an algorithm is T-strongly $\alpha$-competitive if, for any instance of $\ofa$ on a tree metric, the cost incurred by the algorithm, measured by the path distance, is at most $\alpha$ times the optimal offline cost measured by the max-weight distance. Since the max-weight distance is shorter than the path distance, the T-strong competitive ratio is greater than the standard competitive ratio in general.

In this paper, we demonstrate that if a greedy-like algorithm that only uses the positions of a current request and current available server sites (referred to as MPFS~\cite{HIM2023}\footnote{
    MPFS is a class of algorithms that generalize the greedy algorithm (see Definition~\ref{def-mpfs} for details).
})
is designed to be T-strongly $O(k)$-competitive for $\ommt$ with $k$ servers, it can be transformed into an $O(m)$-competitive algorithm for $\ofa(k,m)$ (Theorem~\ref{thm-suffice-for-omt-mpfs}). The proof is completed by establishing the following three claims:
\begin{enumerate}
    \item If there exists a T-strongly $O(k)$-competitive algorithm for $\ommt$ with $k$ servers, then there exists an $O(k)$-competitive algorithm for $\omm(k)$.
    \item If there exists an $O(k)$-competitive algorithm for $\omm(k)$, then there exists an $O(k)$-competitive algorithm for $\ommn(k)$.
    \item If a certain MPFS algorithm is $O(k)$-competitive for $\ommn(k)$, then that algorithm is $O(m)$-competitive for $\ofa(k,m)$.
\end{enumerate}
Claim (2) was previously proven by Meyerson et al.~\cite{MNP2006}, and Claim (3) was proven by Harada et al.~\cite{HIM2023}. In this paper, we prove Claim (1).
To this end, we begin by noting that it suffices to demonstrate a method for obtaining an $O(k)$-competitive algorithm for $\omm(k)$ with a general $k$-point metric space by using a T-strongly $O(k)$-competitive algorithm for $\omm(k)$ on a tree metric (where the edge weights are not necessarily powers of two).

Next, we explain how to convert algorithm $\alg$, which is T-strongly $\alpha$-competitive for a tree metric, into algorithm $\mc{B}$, which is $\alpha$-competitive for a general $k$-point metric space $\mc{M}$. $\mc{B}$ first treats the given metric space as a weighted complete graph, where the weight of each edge corresponds to the distance between its endpoints. $\mc{B}$ then finds a minimum spanning tree (MST) $T$ of this graph and simulates algorithm $\alg$ on $T$.

The key fact to prove that $\mc{B}$ is $\alpha$-competitive for a general metric space $\mc{M}$ is that the distance between any two points in the original metric space $\mc{M}$ lies between the max-weight distance and the path distance on the MST $T$.
Using this fact, we can verify that $\mc{B}$ is $\alpha$-competitive as follows: First, the $\mc{B}$'s cost measured by the distance in $\mc{M}$ is at most the $\alg$'s cost measured by the path distance on $T$. Then, by the T-strong competitiveness of $\alg$, the $\alg$'s cost measured by the path distance is at most $\alpha$ times the optimal offline cost measured by the max-weight distance. Since the optimal offline cost measured by the max-weight distance is not greater than the optimal offline cost measured by the distance in $\mc{M}$, it follows that the $\mc{B}$'s cost is at most $\alpha$ times the optimal offline cost measured by the distance in $\mc{M}$.

Finally, we briefly explain why the distance in the original metric space $\mc{M}$ lies between the max-weight distance and the path distance on the MST $T$.
By the triangle inequality, the distance in $\mc{M}$ is at most the path distance. To see that the distance in $\mc{M}$ is at least the max-weight distance, consider that if the distance in $\mc{M}$ between two points $u$ and $v$ were smaller than the max-weight distance on $T$, one could create a spanning tree with a smaller weight by adding edge $(u,v)$ to $T$ and removing the heaviest edge on the $u$-$v$ path in $T$. This would contradict $T$ being an MST.

The above discussion reduces the task of designing an $O(m)$-competitive algorithm for $\ofa(k,m)$ to designing a T-strongly $O(k)$-competitive MPFS algorithm for $\ommt$ with $k$ servers.

\subsubsection*{Overview of our algorithm}

In the following, we provide an overview of the T-strongly $O(k)$-competitive algorithm for $\ommt(T)$ with $k$ servers, which we refer to as Subtree-Decomposition (SD). Note that in $\ommt$, the set of vertices coincides with the set of servers, meaning that $k = |V(T)| = |E(T)| + 1$.

The proposed algorithm is based on a simple depth-first search (DFS) algorithm. The DFS-based algorithm
begins by arbitrarily selecting one vertex in $T$ as the root before receiving any requests. When a request arrives at a vertex, the algorithm performs a DFS starting from that vertex, and assigns the request to the first available server it encounters.

Intuitively, one can understand why the T-strong competitive ratio of this DFS-based algorithm is at most $O(k)$ when the given tree $T$ is unweighted (where each edge has unit weight) as follows: The nature of DFS ensures that any edge will be traversed at most twice by the algorithm's assignments. In instances where there exist edges traversed more than twice, the optimal offline cost also increases in proportion to the number of such edges, resulting in a sufficiently small ratio of the algorithm's cost to the optimal offline cost\footnote{Note that this statement is imprecise, as discussed and justified in this paper through the analysis via `hybrid algorithm'~\cite{GL2012}.}. Thus, the algorithm's cost measured by the path distance is roughly $O(|E(T)|) = O(k)$. On the other hand, for any request sequence where the algorithm incurs a non-zero cost, the optimal offline cost measured by the max-weight distance is at least 1. Therefore, the ratio of the algorithm's cost measured by the path distance to the optimal offline cost measured by the max-weight distance is $O(k)$.

However, naive application of this algorithm to a weighted tree results in inefficiencies. For instance, if a request occurs at a vertex where the edges connecting it to its children have very large weights, while the edge connecting it to its parent has a very small weight, the algorithm will prioritize assigning the request to a more costly child. To address this inefficiency, SD performs a stepwise DFS, assigning the request at vertex $v$ to the first available server it encounters by prioritizing the exploration of lighter edges. Specifically, SD proceeds as follows:
\begin{enumerate}
    \item Perform a DFS from $v$, restricted to edges with weights no greater than 1.
    \item If no available server is found in step 1, return to $v$ and perform another DFS, this time restricted to edges with weights no greater than 2.
    \item In subsequent steps, double the threshold for edge weights and perform a DFS from $v$ until an available server is found.
\end{enumerate}

Below, we provide an intuitive explanation for why SD is T-strongly $O(k)$-competitive for $\ommt(T)$ with $k$ servers. Let $2^n$ be the weight of the heaviest edge used in the optimal offline assignment. In this case, the optimal offline cost measured by the max-weight distance is at least $2^n$. SD is designed to minimize the number of times it traverses heavy edges, and it can be shown that no edge with weight greater than $2^n$ is traversed in SD's assignment. Furthermore, an edge with weight exactly $2^n$ is traversed at most twice by the nature of DFS. Similarly, edges with weight $2^{n-1}$ are traversed up to two times during the $(n-1)$-th DFS and another two times during the final $n$-th DFS, for a total of four traversals. Extending this reasoning, for each $i = 0, \ldots, n$, the number of traversals for an edge with weight $2^i$ in SD's assignment is expected to be at most $2(n-i+1)\le 2^{n-i+1}$. Therefore, the algorithm's cost measured by the path distance is roughly at most $O(2^i \times 2^{n-i+1} \times |E(T)|) = O(2^n |E(T)|)$, and the ratio to the optimal offline cost measured by the max-weight distance is $O(|E(T)|) = O(k)$.

\subsection{Related Work}
\label{sebsec-related}


\noindent\textbf{Online Metric Matching on a Line.} \quad
A line metric is one of the most interesting and well-studied metric spaces for these problems.
In particular, $\ommn$ on a line has been actively researched~\cite{KN2003,GL2012,ABNPS2014,AFT2018,R2018}. 
The best upper bound on the competitive ratio for $\ommn$ on a line is $O(\log k)$~\cite{R2018},
which is achieved by the Robust-Matching algorithm~\cite{R2016}, and
the best lower bound on the competitive ratio~\cite{PS2021} is $\Omega(\sqrt{\log k})$,
where $k$ denotes the number of servers.
Note that the lower bound $\Omega(\sqrt{\log k})$ can also be applied to any randomized algorithm for $\ommn$ on a line.
As can be seen from the above, the best possible competitive ratio for $\ommn$ on a line has remained open.

There are also some studies on $\ofa$ on a line.
Ahmed et al.~\cite{ARK2020} addressed competitive analysis for $\ofa$ on a line
under the assumption that the server sites are evenly placed and each site has the same capacity.
Under this assumption,
they showed (with rough proofs) that
the natural greedy algorithm is $4m$-competitive 
and the Permutation algorithm (called \textit{Optimal-fill} in their paper) is $m$-competitive for any $m > 2$.
On the other hand,
Harada and Itoh~\cite{HI2023} studied $\ofa$ on a line with a general layout of servers.
They proposed an $(2\alpha(S)+1)$-competitive algorithm called \textit{PTCP} (Policy Transition at Critical Point), where $\alpha(S)$ is the ratio of the diameter of a set $S$ of $m$ server sites to the maximum distance between two adjacent server sites.
They also constructed a layout of servers where PTCP has a constant competitive ratio
while Permutation (or Optimal-Fill) has at least an $\Omega(m)$ competitive ratio.
\half

\noindent\textbf{Online Transportation Problem with a Weaker Adversary.} \quad
For $\ofa$, models with a weaker adversary have also been well analyzed. Here, we will introduce two previous studies.
Kalyanasundaram and Pruhs~\cite{KalP1995} studied
$\ofa$ under the weakened adversary model
where the adversary has only half as many capacities of each server site
as the online algorithm and the length of a request sequence is at most $k/2$.
They showed that the greedy algorithm is
$\Theta(\min (m, \log k))$-competitive and
presented an $O(1)$-competitive algorithm under this model.
Chung et al.~\cite{CKP2008} also studied $\ofa$ under
another weakened adversary where
the adversary has one less capacity of each server site
against the online algorithm.
Under this model,
they presented an $O(\log m)$-competitive deterministic
algorithm on an $\alpha$-HST~\cite{CKP2008} metric
where $\alpha=\Omega(\log m)$
and an $O(\log^3 m)$-competitive randomized algorithm
on a general metric.
\half

\noindent\textbf{Randomized Algorithms for Online Metric Matching.} \quad
There are also many studies on randomized algorithms for $\ommn$.
For a general metric space, Meyerson et al.~\cite{MNP2006} showed the lower bound $\Omega(\log k)$ and the upper bound $O(\log^3 k)$ on the competitive ratio for $\ommn$ with $k$ servers.
The upper bound was improved to be $O(\log^2 k)$ by Bansal et al.~\cite{BBGN2007}.
As can be seen from the above, there still has been a gap between the upper bound $O(\log^2 k)$ and the lower bound $\Omega(\log k)$ for $\ommn$.
For doubling metrics, Gupta and Lewi~\cite{GL2012} showed an $O(\log k)$-competitive randomized algorithm.

Randomized algorithms for $\ofa(k,m)$ have not been studied for a long time, but recently Kalyanasundaram et al.~\cite{kalyanasundaram2023randomized} proposed an $O(\log^2 m)$-competitive algorithm.
\half

\noindent\textbf{Stochastic Online Metric Matching.} \quad
Raghvendra \cite{R2016} considered $\ommn$ under the random arrival model in which the adversary chooses the set of request locations at the start but the arrival order is a permutation chosen uniformly at random from the set of all possible permutations.
He proposed an algorithm called \textit{Robust-Matching} and showed that it is $(2H_k-1)$-competitive and best possible for $\ommn(k)$ under the random arrival model,
where $H_k$ denotes the $k$-th harmonic number.
Furthermore, Robust-Matching also has the best possible competitive ratio of $2k-1$ under the normal adversarial model.

Gupta et al.~\cite{gupta2019stochastic} considered $\ommn$ under online i.i.d. arrivals.
In this model, requests are drawn independently from a known probability distribution over a metric space.
They proposed an algorithm called \textit{FAIR-BIAS} and showed that it is $O((\log\log\log k)^2)$-competitive for a general metric space and $9$-competitive for a tree metric under this model. 

\section{Preliminaries}
\label{sec-preliminary}

\subsection{Definition of Problems}
\label{subsec-def-ofa}

In this section, we define the online metric matching, the online transportation and their variants.

\subsubsection*{Online Metric Matching}
\label{subsubsec-def-omm}

To begin with, we define the online metric matching (denoted by $\ommn$).
An instance of $\ommn$ is represented by a quadruple $I=(X,d,S,\sigma)$, where
$X$ is a set of points, $d:X\times X\to\mathbb{R}_{\geq 0}$ is a distance function on $X$,
$S$ is a finite subset of $X$, and
$\sigma=r_1\ldots r_{|S|}\in X^{|S|}$.
An element of $S$ is called a \textit{server} and the $t$-th entry $r_t$ of $\sigma$
is called the \textit{request at time $t$}
or \textit{$t$-th request}.

$X$, $d$ and $S$ are given to an online algorithm in advance,
while requests are given one-by-one from $r_1$ to $r_{|S|}$.
At each step, an online algorithm $\alg$ for $\ommn$ maintains `assignment'
that is initialized to $\emptyset$. 
When a request $r_{t}$ is revealed, 
$\alg$ must assign $r_{t}$ to one of the available servers irrevocably.
If $r_{t}$ is assigned to the server $s_{t}$, then the pair $(r_{t},s_{t})$
is added to the current assignment and the cost $d(r_{t},s_{t})$ is incurred for this pair.
The cost of the assignment is the sum of the costs of all the pairs contained in it.
The goal of an online algorithm is to minimize the cost of the final assignment.

We use the following notation on $\ommn$.

\begin{enumerate}
\item $\ommn(k)$ denotes $\ommn$ with $k$ servers.
\item $\omm$ denotes $\ommn$ where requests arrive at the position of some server, i.e., $X=S$.
We simply write an instance $I$ of $\omm$ as $I=(S,d,\sigma)$ instead of $(S,d,S,\sigma)$. 
\item $\omm(k)$ denotes $\omm$ with $k$ servers.
\end{enumerate}

\subsubsection*{Two Metrics on Edge-Weighted Trees}

Before presenting online metric matching on a power-of-two tree metric, we first introduce the definitions and notations for the two distance functions used for edge-weighted trees.
For an edge-weighted tree $T$, we use $d_T$ to denote the path distance on $V(T)$, i.e.,
for any $u, v\in V(T)$,
$
    d_T(u,v)\coloneqq \sum_{e\in E(P_T(u,v))}w_T(e),
$
where $P_T(u,v)$ denotes the unique simple path in $T$ from $u$ to $v$.
In this paper, we define another distance $d_T^{\max}$ on $T$ as follows:
\[
    d_T^{\max}(u,v)
    \coloneqq \max_{e\in E(P_T(u,v))}w_T(e).
\]
We call $d_T^{\max}$ the \textit{max-weight distance} on $T$.
It is easy to verify that $d_T^{\max}$ defines a metric on $V(T)$ when each edge has a positive weight.
By the above definition, we can observe the following relationship between the path distance and the max-weight distance.

\begin{remark}
\label{rem-two-distance-k-1}
For any edge-weighted tree $T$ and any $u, v\in V(T)$,
we have
\[
d_T^{\max}(u,v) \le d_T(u,v) \le |E(T)| \cdot d_T^{\max}(u,v),
\]
where $d_T$ denotes the path distance on $T$ and $d_T^{\max}$ denotes the max-weight distance on $T$.
\end{remark}

\subsubsection*{Online Matching on a Power-of-two Tree Metric}

Then, we introduce a notion of power-of-two weighted tree and define a special case of $\omm$ called online matching on a power-of-two tree metric (denoted by $\ommt$).

\begin{definition}[Power-of-two Weighted Tree]
\label{def-po2wt}
Let $T=(V(T),E(T))$ be a tree and $w_T:E(T)\to \mathbb{R}_{\geq 0}$ be a weight function of $T$. We say that $T$ is a \textsf{power-of-two weighted tree} if
for any $e\in E(T)$, there exists a non-negative integer $i$ such that $w_T(e)=2^i$.
\end{definition}

\begin{definition}[Power-of-two Tree Metric]
\label{def-po2tm}
We say that a metric space $(X,d)$ is a \textsf{power-of-two tree metric} if there exists a power-of-two weighted tree $T$ such that $V(T)=X$ and $d=d_T$. In this case, we also say that $(X,d)$ is \textsf{induced} by $T$.
\end{definition}

The online matching on a power-of-two tree metric is
a variant of the online metric matching problem
where a metric space $(X,d)$ is induced by a power-of-two weighted tree $T$ and
a set $S$ of servers is $V(T)$.
In other words, an instance of $\ommt$ is
an instance $I=(X,d,S,\sigma)$ of $\ommn$ such that $X=S=V(T)$ and $d=d_T$.

For a power-of-two weighted tree $T$, $\ommt(T)$ denotes $\ommt$ where the metric space is induced by $T$. We simply write an instance $I$ of $\ommt(T)$ as $I=(T,\sigma)$ instead of $(V(T),d_T,V(T),\sigma)$ for $\ommn$.

\subsubsection*{Online Transportation Problem}

Finally, we define the online transportation problem (denoted by $\ofa$).
In this problem, $k$ servers are clustered
in specific $m$ $(\leq k)$ locations
called \textit{server sites}.
The number of servers at a server site is called the \textit{capacity} of that site.
Assume that each site has a capacity of at least 1. 
An instance of $\ofa$ is represented by a quintuple $(X, d, S, c,\sigma)$.
Here, $(X, d)$ represents the metric space,
$S$ represents the set of server sites,
$c:S\to\mathbb{N}$ represents the capacity of each site, and
$\sigma=r_1\ldots r_{k}\in X^{k}$ represents the request sequence\footnote{
For $\ofa$, the number of requests is generally set to be \underline{at most} $k$.
As can be seen from~\cite[Lemma 2.1]{HIM2023}, however, the assumption that the number of requests is \underline{exactly} $k$ has no effect on the competitive analysis.
}.
$\ofa$ is the same as $\ommn$ except that an online algorithm can assign up to $c(s)$ requests to one server site $s$.
In other words, $\ommn$ can be viewed as a special case of $\ofa$ where $k=m$ and each site has unit capacity.
We use $\ofa(k,m)$ to denote $\ofa$ with $k$ servers and $m$ $(\leq k)$ server sites.

\subsection{Notation and Terminology}
\label{subsec-notation}
Let $I=(X,d,S,c,\sigma)$ be any instance of $\ofa(k,m)$, $r_t$ be the $t$-th request of $I$ and $\alg$ be any online algorithm for $\ofa$.
We use $\svrinst{t}{I}{\alg}$ to denote the server to which $\alg$ assigns $r_t$ when processing $I$. Let $\alg(I)$ be the total cost incurred when $\alg$ processes $I$, i.e.,
\[
\alg(I)=\sum_{t=1}^{k}d(r_t,\svrinst{t}{I}{\alg}).
\]
$\opt$ denotes the optimal offline algorithm, i.e., $\opt$ knows the entire $\sigma$ in advance and assigns $r_t$ to $\svrinst{t}{I}{\opt}$ to minimize the total cost $\opt(I)$.
At any step of the execution of an online algorithm, a server site is called \textit{free}
if the number of requests assigned to it is less than its capacity, and \textit{full} otherwise.
Let $F_t(\alg,I)$ be the set of all free server sites just after $\alg$ assigns $r_t$ to a server.
We say that $\alg$ is $\alpha$-competitive if $\alg(I)\leq \alpha\cdot\opt(I)$ for any instance $I$ of $\ofa$. The competitive ratio $\mc{R}(\alg)$ of $\alg$ is defined to be the infimum of $\alpha$ such that $\alg$ is $\alpha$-competitive, i.e., $\mc{R}(\alg) = \inf\{\alpha : \text{$\alg$ is $\alpha$-competitive}\}$.
In this paper, we consider only algorithms that,
when a request $r$ arrives at the position of a free server site, assign $r$ to that site\footnote{
Algorithms that do not satisfy this condition are known to be not $\alpha$-competitive for any $\alpha>0$.
}.

For an instance $I=(X,d,S,\sigma)$ of $\ommn(k)$, we simply say that a server is \textit{free} when no request is assigned to it, and \textit{full} otherwise.
In other respects, we use the same notation and terminology as in $\ofa$.

\subsection{Technical Lemmas from Prior Work}
\label{subsec-tech-lemma}

In this section, we present two prior results that play crucial roles in achieving our objective of designing an $O(m)$-competitive deterministic algorithm for $\ofa(k, m)$.
The following theorem claims that if there exists an $O(k)$-competitive algorithm for $\omm(k)$, then there also exists an $O(k)$-competitive algorithm for $\ommn(k)$.
\begin{theorem}[Meyerson et al.~\cite{MNP2006}]
\label{thm-server-eq}
If there exists an $\alpha$-competitive algorithm $\alg$ for $\omm$, then
there exists a $(2\alpha +1)$-competitive algorithm $\mc{B}$ for $\ommn$.
\end{theorem}
The algorithm $\mc{B}$ is designed as follows: When a request arrives, it is first moved to the nearest server site (not necessarily available), and then assigned to a server according to $\alg$.


Next, we introduce a class of algorithms for $\ofa$ called MPFS (Most Preferred Free Servers) and remark that 
to design an $O(m)$-competitive algorithm for $\ofa(k,m)$,
it suffices to design an $O(k)$-competitive MPFS algorithm for $\omm(k)$.

\begin{definition}[MPFS Algorithm~\cite{HIM2023}]
\label{def-mpfs}
Let $\alg$ be a deterministic online algorithm for $\ofa$.
We say that 
$\alg$ is an \textsf{MPFS (most preferred free servers)} algorithm if it deals with a request sequence $\sigma=r_1\ldots r_{k}$ as follows:
\begin{enumerate}
\item For each $1 \leq t \leq k$, 
the priority (with no ties) of all server sites for $r_{t}$ is determined by only the positions of $r_{t}$ and all server sites $S$,
\item $\alg$ assigns $r_{t}$ to a server with the highest priority among all free server sites $F_{t-1}(\alg,I)$.
\end{enumerate}
\end{definition}

\noindent
Let $\mc{MPFS}$ be the class of MPFS algorithms for $\ofa$.
By its definition, given a new request $r$ and
a set $F\subseteq S$ of current free server sites, an MPFS algorithm $\alg$ uniquely determines a server $s$ to which $r$ is assigned.
We use $\svrmpfs{r}{F}{\alg}$ to denote such $s$, i.e.,
a server to which $\alg$ assigns $r$ when $F$ is a set of free server sites.
For any MPFS algorithm, it is immediate that the following remark holds.

\begin{remark}
\label{rem-mpfs}
Let $\alg\in\mc{MPFS}$ and $s=\svrmpfs{r}{F}{\alg}$.
If $s\in F' \subseteq F$, then $\svrmpfs{r}{F'}{\alg}=s$.
\end{remark}

\noindent
Moreover, the following strong theorem~\cite{HIM2023} is known for MPFS algorithms.

\begin{theorem}[Harada et al.~\mbox{\cite[Corollary 3.10]{HIM2023}}]
\label{thm-mpfs}
Let $\alg\in \mc{MPFS}$ and suppose that 
$\alg$ is $\alpha(k)$-competitive for $\ommn(k)$, where $\alpha(k)$ is a non-decreasing function of $k$.
Then, for any $m \le k$, $\alg$ is $\alpha(m)$-competitive for $\ofa(k,m)$.
\end{theorem} 
\begin{proof}[Proof Sketch]
For simplicity, consider an instance $I=(X,d,S,c,\sigma)$ where $c(s)\equiv\ell>1$, i.e., the number of requests and servers is $ k = m\ell$.  
By Hall's theorem, we can `partition' the request sequence $ \sigma = r_1 \ldots r_k $ into $\ell$ request sequences $\sigma_i = r_1^i \ldots r_m^i$ for $i = 1,\ldots,\ell$, such that both $\alg$ and $\opt$ assign requests in $\sigma_i$ to distinct $m $ server sites.  

By the definition of the MPFS algorithm, when executing the instance $I_i = (X,d,S,\sigma_i)$ for $\ommn$ (i.e., when all server sites have unit capacity), each request is assigned to the same server site as in the original execution of $I$. Thus, we have $\alg(I) = \sum_{i} \alg(I_i)$. Similarly, by optimality, we also have $\opt(I)=\sum_{i} \opt(I_i)$.
Therefore, $\max_i \alg(I_i)/\opt(I_i)$ is an upper bound on the competitive ratio of $\alg$. This implies that the competitive ratio of $\alg$ for $\ommn$ dominates the competitive ratio for $\ofa$.  
\end{proof}

By Theorems~\ref{thm-server-eq} and~\ref{thm-mpfs},
if there exists an $\alpha(k)$-competitive MPFS algorithm for $\omm(k)$, then
we easily obtain a $(2\alpha(m)+1)$-competitive algorithm for $\ofa(k,m)$.
Note that if $\alg$ is an MPFS algorithm, then the algorithm $\mc{B}$ shown in Theorem~\ref{thm-server-eq} is also an MPFS algorithm.

\section{T-strong competitive ratio for Tree Metrics}
\label{sec-T-strong-competitive}

In this section, we introduce a new concept of T-strong competitive ratio for $\ommt$ to represent the performance of an algorithm and show that T-strong competitiveness for $\ommt$ is closely related to standard competitiveness for $\omm$.
The most important result in this section is the following theorem.
Thanks to this theorem, our goal of designing an $O(m)$-competitive algorithm for $\ofa(k,m)$ is reduced to designing a T-strongly $O(k)$-competitive MPFS algorithm for $\ommt$ with $k$ servers.

\begin{theorem}
\label{thm-suffice-for-omt-mpfs}
If there exists a T-strongly $\alpha(k)$-competitive MPFS algorithm for $\ommt$ with $k$ servers, then there exists a $(4\alpha(m)+1)$-competitive algorithm $\mc{B}$ for $\ofa(k,m)$, where $\alpha(\cdot)$ is a non-decreasing function. 
\end{theorem}

We begin with defining T-strong competitiveness.
Hereafter, we use $\opt_T^{\max}$ to denote the optimal offline algorithm where the cost of assigning a request to a server is measured by the max-weight distance on $T$.

\begin{definition}[T-strong competitive ratio]
\label{def-t-strongly-comp}
Let $I=(T,\sigma)$ be any instance of $\ommt$ and $\opt_T^{\max}(I)$ denote the minimum cost of assigning all requests to servers, where the cost is measured by the max-weight distance $d_T^{\max}$ on $T$.
We say that an algorithm $\alg$ is \textsf{T-strongly $\alpha$-competitive} if, for any instance $I$ of $\ommt$, it follows that
\[
\alg(I)\le \alpha\cdot\opt_T^{\max}(I),
\]
where the cost $\alg(I)$ is measured by the path distance $d_T$ on $T$. 
In addition, the \textsf{T-strong competitive ratio} of $\alg$ is defined to be the infimum of $\alpha$ such that $\alg$ is T-strongly $\alpha$-competitive.
\end{definition}

\begin{remark}
\label{rem-alg-cost-notation}
In the rest of this paper, the cost of an algorithm is generally measured by the path distance, unless we specifically use the notations $\opt_T^{\max}$ or $d_T^{\max}$.
\end{remark}

\begin{remark}
\label{rem-t-strong-is-stronger}
By Remark~\ref{rem-two-distance-k-1}, we have $\opt_T^{\max}(I)\le \opt(I)$. Therefore, if an algorithm $\alg$ is T-strongly $\alpha$-competitive for $\ommt(T)$, then $\alg$ is also $\alpha$-competitive for $\ommt(T)$.
\end{remark}

The following theorem presents a method to convert a T-strongly $\alpha$-competitive algorithm for $\ommt$ into a $2\alpha$-competitive algorithm for $\omm$ with a general metric.

\begin{theorem}
\label{thm-storongly-competitive-to-genera-lmetrics}
    If there exists a T-strongly $\alpha$-competitive algorithm $\alg$ for $\ommt$, then there exists a $2\alpha$-competitive algorithm $\mc{B}$ for $\omm$.
\end{theorem}

\begin{proof}
Let $(S,d,\sigma)$ be an instance for $\omm(k)$,
where $|S|=k$.
Without loss of generality, suppose the minimum distance between two distinct servers is 1, i.e.,
$\min_{s,s'\in S, s\neq s'}d(s,s')=1$.

$\mc{B}$ first constructs a power-of-two weighted tree $T$ by using $(S,d)$ and
then simulates $\alg$ with input instance $I'=(T,\sigma)$.
Now we describe how to construct $T$ by using $(S,d)$.
First, we represent the $k$-point metric space $(S, d)$ as an edge-weighted $k$-vertex complete graph $K_{S,d}$ in which each edge $(u, v)$ has a weight $d(u, v)$. Let $T'$ be a minimum spanning tree of the $K_{S,d}$.
Next, we obtain a power-of-two weighted tree $T$ by adjusting the edge weights of the $T'$ as follows: for each edge $e\in E(T')$, if $2^{i-1}<w_{T'}(e)\leq 2^i$, then $w_{T}(e)\coloneqq2^i$, i.e., $w_{T}(e)=2^{\lceil\log_2 w_{T'}(e)\rceil}$.

To prove that $\mc{B}$ is $2\alpha$-competitive, we use the following claim which establishes the relationships between the two types of metrics on $T$ and the original metric.

\begin{claim}
\label{claim-sd-dist}
Let $(S,d,\sigma)$ be any instance for $\omm(k)$ and $T$ be a power-of-two weighted tree constructed by $\mc{B}$. Then, for any two different points $u,v\in S$,
\[
d_T^{\max}(u,v)< 2d(u,v) \leq 2d_T(u,v).
\]
\end{claim}

\begin{claimproof}
The second inequality is trivial by the triangle inequality.
Thus, we show the first inequality.
By the definition of $T$ and $T'$, $w_T(e)< 2w_{T'}(e)$ holds for any edge $e\in E(T)=E(T')$. Therefore, it suffices to show that $d_{T'}^{\max}(u,v)\leq d(u,v)$ for any $u,v\in S$.

By contradiction,
assume that there exist two vertices $u,v\in S$ such that
$d_{T'}^{\max}(u,v)> d(u,v)$.
This implies that there exists an edge $(u',v')\in P_{T'}(u,v)$ such that $d(u',v')=w_{T'}(u',v')>d(u,v)$. However, this contradicts the fact that $T'$ is a minimum spanning tree of $K_{S,d}$ since we have a spanning tree with less weight by removing edge $(u',v')$ from $T'$ and adding edge $(u,v)$ to $T'$. Thus, we have $d_{T'}^{\max}(u,v)\leq d(u,v)$.
\end{claimproof}

Then, we can prove $\mc{B}$ to be $2\alpha$-competitive as follows:
\begin{align*}
\mc{B}(I)
& = \sum_{t=1}^k d(r_t,\svrinst{t}{\alg}{I'})
\le \sum_{t=1}^k d_T(r_t,\svrinst{t}{\alg}{I'})
\le \alpha \cdot \sum_{t=1}^k d_T^{\max}(r_t,\svrinst{t}{\opt_T^{\max}}{I'}) \\
& \le \alpha \cdot \sum_{t=1}^k d_T^{\max}(r_t,\svrinst{t}{\opt}{I})
\le 2\alpha \cdot \sum_{t=1}^k d(r_t,\svrinst{t}{\opt}{I})
= 2\alpha \cdot \opt(I),
\end{align*}
where the first and last inequality is due to Claim~\ref{claim-sd-dist}, the second inequality holds by T-strong $\alpha$-competitiveness of $\alg$, and the third inequality holds by the optimality of $\opt_T^{\max}$ for $I'$.
\end{proof}

By Theorems~\ref{thm-storongly-competitive-to-genera-lmetrics}, \ref{thm-server-eq} and~\ref{thm-mpfs}, we obtain Theorem~\ref{thm-suffice-for-omt-mpfs}.
Thus, in the rest of this paper, we will focus on designing an MPFS algorithm for $\ommt$.

%
%

%
%
%
%
\section{New Algorithm: Subtree-Decomposition}
\label{sec-our-algorithm}

In this section, we propose a new MPFS algorithm for $\ommt$ called Subtree-Decomposition (SD).
In the subsequent sections,
we use $\algs$ to denote the SD algorithm
unless otherwise specified.

%
%
%
%
%
\subsection{Notation for Graphs and Trees}
\label{subsec-note-tree}

To begin with, we describe the notation for graphs and trees used in this paper.
The following notation is used for any two graphs $G=(V(G), E(G))$ and $G'=(V(G'),E(G'))$.

\begin{itemize}

\item $G \cap G'$ denotes the graph $(V(G)\cap V(G'),E(G)\cap E(G'))$.

\item $G \cup G'$ denotes the graph $(V(G)\cup V(G'),E(G)\cup E(G'))$.

\item $G\setminus G'$ denotes the minimal subgraph of $G$ that contains all edges in $E(G)\setminus E(G')$.

\end{itemize}

Let $T=(V(T),E(T))$ be a power-of-two weighted tree and $d_T$ be the path distance on $T$. Suppose that $T$ is rooted at $\rho\in V(T)$.
For $T$ and any subtree $U$ of $T$, we use the following notation.

\begin{itemize}

\item For $v\in V(T)$, $\parent(v)$ denotes the parent of $v$.

\item For $u,v \in V(T)$, $\lca(u,v)$ denotes the least common ancestor of $u$ and $v$.

\item $\rho(U)$ denotes the closest vertex in $V(U)$ to the root $\rho$. We simply refer to $\rho(U)$ as the \textit{root} of $U$.

\item $w_U^{\max}$ denotes the maximum weight in $U$, i.e., $w_U^{\max}\coloneqq\max_{(u,v)\in E(U)}d_T(u,v)$.

\item $E_{\max}(U)$ denotes the set of heaviest edges in $U$, i.e., \[E_{\max}(U)\coloneqq\{ (u,v)\in E(U): d_T(u,v)=w_U^{\max} \}.\]

\end{itemize}

The notations $v\in V(T)$ and $e\in E(T)$ are sometimes abbreviated to $v\in T$ and $e\in T$ respectively
when context makes it clear.

\subsection{Definition of Subtree-Decomposition}
\label{subsec-def-alg}

Then, we define the SD algorithm denoted by $\algs$.
Before presenting the formal definition, we describe the intuitive behavior of SD. 

\half
\noindent\textbf{Intuitive behavior.} \quad
Let $T=(V(T),E(T))$ be a given power-of-two weighted tree rooted at an arbitrary vertex $\rho \in V(T)$.
For each vertex $v$, let $W_i(v)\subseteq V(T)$ be the set of vertices reachable from vertex $v$ only through edges with weights at most $2^i$. When a new request arrives at vertex $v$, for $i=1,2,\ldots$, SD performs a DFS starting from $v$ to search for a free server in $W_i(v)$. SD then assigns the request to the first free server located by a DFS.



\half
\noindent\textbf{Decomposition of a Tree.} \quad
Next, to describe the algorithm more precisely and simplify the inductive analysis, we redefine the above algorithm recursively. To this end, we decompose a given power-of-two weighted tree $T=(V(T),E(T))$ into subtrees (hence the algorithm's name) as follows:

Let $\rho^{(1)}\coloneqq\rho$ and 
we define subtrees $T^{(1)}$ and $T^{(2)}$
as follows:
We arbitrarily choose $\rho^{(2)}$,
one of the children of the root $\rho$ $(=\rho^{(1)})$,
and consider a graph
$T\setminus(\rho^{(1)},\rho^{(2)})$ obtained by removing edge $(\rho^{(1)},\rho^{(2)})$ from $T$.
Note that
$T\setminus(\rho^{(1)},\rho^{(2)})$ consists of
two subtrees.
Let $T^{(1)}$ be the subtree that contains $\rho^{(1)}$ and $T^{(2)}$ be the other subtree that contains $\rho^{(2)}$.

Next, let $\rho_0\coloneqq\rho$ and we define subtrees $T_0,T_1,\ldots$ as follows:
Let $T_0$ be the maximal subtree of $T$ that contains the root $\rho$ $(=\rho_0)$ and does not contain any heaviest edge in $E_{\max}(T)$.
We use $\rho_1,\ldots ,\rho_l$ to denote the vertices $v$ such that
$v\notin T_0$ and $\parent(v)\in T_0$.
Note that $(\rho_i, \parent(\rho_i)) \in E_{\max}(T)$ by the definition of $T_0$.
For $i=1,\ldots,l$, let $T_i$ be the subtree of $T$ rooted at $\rho_i$.
By the definition of $\{T_i\}_{i=0}^l$, $\{V(T_i)\}_{i=0}^l$ is a partition of $V(T)$.
Then, for any vertex $v$, there uniquely exists a subtree $T_i$ such that $v\in T_i$.
Let $T_{-i}$ denote the subtree of $T$ induced by $V(T)\setminus V(T_i)$.

\half
\noindent\textbf{Recursive Definition.} \quad
Now we are ready to describe the formal and recursive definition of SD.
For the base case where $|V(T)|=1$, SD assigns each request to the unique server.
For $j=1,2$, let $\algs_{(j)}$ be the SD algorithm for $T^{(j)}$ rooted at $\rho^{(j)}$ and
for $i=0,\ldots,l$, let $\algs_i$ be the SD algorithm for $T_i$ rooted at $\rho_i$.
$\algs$ has two phases:
When all servers in $T_0$ are full,
Phase 1 terminates and Phase 2 follows.

Let $r$ be a new request and $F\subseteq V(T)$ be a set of free servers. $\algs$ assigns $r$ by using $\algs_{(j)}$ for $j=1,2$
and $\algs_i$ for $i=1,\ldots,l$.

\half
\noindent \textbf{Phase 1: There is at least one free server in \boldmath{$T_0$}.}

Assume that $r\in T_i$ for some $i=0,\ldots,l$.
If there exists a free server in $T_i$, then $\algs$ assigns $r$ to a server in $T_i$ according to $\algs_i$.
Otherwise,
$\algs$ assigns $r$ to a server in $T_0$
according to $\algs_0$
by regarding
$\parent(\rho_i)$
as a new request and
$F\cap V(T_0)$ as a set of free servers.


\half
\noindent \textbf{Phase 2: There is no free server in \boldmath{$T_0$}.}

Assume that $r\in T^{(j)}$ for some $j=1,2$ and $r\in T_i$ for some $i=0,\ldots,l$.
If there exists a free server in $T_i$, then $\algs$ assigns $r$ to a server in $T_i$ according to $\algs_i$.
Otherwise, if there exists a free server in $T^{(j)}$, then $\algs$ assigns $r$ to a server in $T^{(j)}$ according to $\algs_{(j)}$.
Otherwise, $\algs$ assigns $r$ to a server in $T^{(3-j)}$
according to $\algs_{(3-j)}$
by regarding
$\rho^{(3-j)}$
as a new request and
$F\cap V(T^{(3-j)})$
as a set of free servers.
\half

%
%
%
%

\begin{figure}[htbp]
\begin{algorithm}[H]
\caption{Subtree-Decomposition (denoted by $\algs$)}
\label{alg-sdalg}
\begin{algorithmic}
\REQUIRE A power-of-two weighted tree $T$, a request $r$ and a set $F\subseteq V(T)$ of free servers.
\STATE Suppose that $r\in T_i$ and $r\in T^{(j)}$ for some $i=0,\ldots,l$ and $j=1,2$.
\IF{$\algs$ is in Phase 1, i.e., $F \cap V(T_0) \neq \emptyset$}
\IF{there exists a free server in $T_i$}
\STATE Assign $r$ to a server in $T_i$ according to $\algs_i$.
\ELSE
\STATE Assign $r$ to a server in $T_0$
according to $\algs_0$ by regarding $\parent(\rho_i)$ as a new request and $F\cap V(T_0)$ as a set of free servers.
\ENDIF
\ENDIF
\IF{$\algs$ is in Phase 2, i.e., $F \cap V(T_0)=\emptyset$}
\IF{there exists a free server in $T_i$}
\STATE Assign $r$ to a server in $T_i$ according to $\algs_i$.
\ELSIF{there exists a free server in $T^{(j)}$}
\STATE Assign $r$ to a server in $T^{(j)}$ according to $\algs_{(j)}$.
\ELSE
\STATE Assign $r$ to a server in $T^{(3-j)}$
according to $\algs_{(3-j)}$
by regarding $\parent(\rho^{(3-j)})$ as a new request and $F\cap V(T^{(3-j)})$ as a set of free servers.
\ENDIF
\ENDIF
\end{algorithmic}
\end{algorithm}
\end{figure}

We establish that SD is an MPFS algorithm through the following proposition.
\begin{proposition}
\label{prop-sd-is-mpfs}
    Subtree-Decomposition defined in Algorithm~\ref{alg-sdalg}
    is an MPFS algorithm.
\end{proposition}

\begin{proof}
The proof is by induction on $|V(T)|$.
For the base case $|V(T)|=1$,
all algorithms assign a request to the unique server and
it is immediate that $\algs\in\mc{MPFS}$.
For the inductive step,
consider a power-of-two weighted tree $T$ with $k$ vertices
and suppose that
$\algs \in \mc{MPFS}$
for any power-of-two weighted tree with less than $k$ vertices.
By this induction hypothesis,
$\algs_i$ for $i=0,\ldots,l$ and $\algs_{(j)}$ for $j=1,2$
are in $\mc{MPFS}$.

For any $\alg\in\mc{MPFS}$ for $\ommt(T)$,
any request $r\in V(T)$ and
any set $F\subseteq V(T)$ of free servers,
let
$\priority{\alg}{r}{F}\in V(T)^{|F|}$
be the list of elements of $F$
sorted in order of $\alg$'s priority for $r$.
We refer to this list as
the preference list of $\alg$ for $r$ with respect to $F$.
For any two lists $L=(v_1,\ldots,v_n)$ and $L'=(v'_1,\ldots,v'_m)$,
we use $L\circ L'$ to denote the concatenation of $L$ and $L'$, i.e.,
$L\circ L'=(v_1,\ldots,v_n,v'_1,\ldots,v'_m)$.

To prove that $\algs\in\mc{MPFS}$,
it suffices to show that, for any $r\in V(T)$,
$\priority{\algs}{r}{V(T)}$ is determined only by
$r$ and $T$.
If $r\in V(T_0)\cap V(T^{(j)})$ for some $j=1,2$, then, by its definition,
$\algs$ first tries to assign $r$ to a server in $V(T_0)$ according to $\algs_0$,
then tries to assign to a server in $V(T^{(j)})\setminus V(T_0)$ according to $\algs_{(j)}$
and finally assign to a server in $V(T^{(3-j)}) \setminus V(T_0)$ according to $\algs_{(3-j)}$
by regarding $r$ as $\rho^{(3-j)}$.
Thus, the preference list for $r$ is
\begin{align*}
    \priority{\algs}{r}{V(T)}
    &= \priority{\algs_0}{r}{V(T_0)}
        \:\circ\: \priority{\algs_{(j)}}{r}{V(T^{(j)}) \setminus V(T_0)} \\
    &   \qquad \circ\: \priority{\algs_{(3-j)}}{\rho^{(3-j)}}{V(T^{(3-j)}) \setminus V(T_0)}.
\end{align*}
By the induction hypothesis,
$\algs_0$, $\algs_{(1)}$ and $\algs_{(2)}$ are MPFS algorithms.
Therefore, $\priority{\algs}{r}{V(T)}$ depends only on $r$ and $T$.

If $r\in V(T_i) \cap V(T^{(j)})$ for some $i=1,\ldots,l$ and $j=1,2$,
then we can similarly see that
\begin{align*}
    \priority{\algs}{r}{V(T)}
    &= \priority{\algs_i}{r}{V(T_i)}
        \:\circ\:\priority{\algs_0}{\parent(\rho_i)}{V(T_0)} \\
    &    \qquad\circ\: \priority{\algs_{(j)}}{r}{V(T^{(j)}) \setminus (V(T_i)\cup V(T_0))} \\
    &    \qquad\circ\: \priority{\algs_{(3-j)}}{\rho^{(3-j)}}{V(T^{(3-j)}) \setminus V(T_0)}.
\end{align*}
Since $\algs_i$, $\algs_0$, $\algs_{(1)}$ and $\algs_{(2)}$ are MPFS algorithms,
$\priority{\algs}{r}{V(T)}$ depends only on $r$ and $T$.
Hence, $\algs$ for $T$ is also an MPFS algorithm.
\end{proof}

Finally, we show that the processing time of SD for each request is $O(m)$.
\begin{proposition}
\label{prop-sd-m-time}
    Subtree-Decomposition processes each request in $O(m)=O(|V(T)|)$ time.
\end{proposition}
\begin{proof}
    The proof is by induction on $|V(T)|$.
    For the base case $|V(T)|=1$, each request is obviously processed in $O(1)$ time. 
    For the inductive step, suppose that the processing time of $\algs_i$ per request is $O(|V(T_i)|)$ for $i=0,\ldots,l$ and that of $\algs_{(j)}$ is $O(|V(T^{(j)})|)$ for $j=1,2$. If SD is in Phase 1 and a request $r$ is in $T_0$, then the processing time is $O(|V(T_0)|)$ by the induction hypothesis.
    If SD is in Phase 1 and a request $r$ is in $T_i$ for some $i=1,\ldots,l$, then the processing time is $O(|V(T_0)|) + O(|V(T_i)|) = O(|V(T)|)$ by the induction hypothesis. Otherwise, i.e., SD is in Phase 2, the processing time is $O(|V(T^{(1)})|) + O(|V(T^{(2)})|) = O(|V(T)|)$ by the induction hypothesis.
    Thus, the proposition holds.
\end{proof}
%
%
%
%
%


\section{Hybrid Algorithm}
\label{sec-hybrid-algorithm}

In this section, we introduce the notion of hybrid algorithms and their properties.
This idea was initiated by Gupta and Lewi~\cite{GL2012}
and very useful in analyzing the competitive ratio of an MPFS algorithm.
Note that all definitions and results in this section are applicable
to any (not necessarily SD) MPFS algorithm.
To begin with, we define hybrid algorithms.

%
%
\begin{definition}[Hybrid Algorithm] \label{def-hybrid}
Let $\alg\in\mc{MPFS}$.
For a positive integer $\start$ and a server $\stsvr$,
consider an algorithm $\hyb=(\alg, \start, \stsvr)$ that
assigns requests for $\omm$ as follows: 
\begin{enumerate}
\item The first  $\start -1$ requests are assigned according to $\alg$,
\item if $\stsvr$ is free just before the $\start$-th request is revealed, then the $\start$-th request is assigned to $\stsvr$ and
the subsequent requests are assigned according to $\alg$, and
\item if $\stsvr$ is full just before the $\start$-th request is revealed, then the $\start$-th and subsequent requests are assigned according to $\alg$.
\end{enumerate}
We call $\hyb=(\alg,\start,\stsvr)$ a \textsf{hybrid algorithm} of $\alg$
with \textsf{{\stime}} $\start$ and \textsf{{\textdecsvr}} $\stsvr$. 
\end{definition}

One can observe that $\alg$ and $\hyb$ output different assignments when decoupling server $\stsvr$ is free at decoupling time $\start$ and $\alg$ assigns the $\start$-th request to a server other than $\stsvr$, i.e., $\stsvr\in F_{\start}(\alg,I)$.
The purpose of considering the hybrid algorithm is to reduce the evaluation of the competitive ratio of a certain MPFS algorithm $\alg$ to evaluating the difference in cost between $\alg$ and its hybrid algorithm $\hyb$. Therefore, we are only interested in cases where the original MPFS algorithm $\alg$ and its hybrid algorithm $\hyb$ output different assignments.
Motivated by this insight, we give the following definition of hybrid instances.

%
%
\begin{definition}[Hybrid Instance] \label{def-hybinst}
Let $I=(S,d,\sigma)$ be any instance of $\omm$ and
$\hyb=(\alg,\start,\stsvr)$ be a hybrid algorithm of $\alg\in\mc{MPFS}$.
We say that $I$ is \textsf{valid} with respect to $\hyb$ if
$\alg$ and $\hyb$ output different assignments when processing $I$, i.e.,
$\stsvr\in F_{\start}(\alg,I)$ and \textsf{invalid} otherwise.
We refer to a pair $H=(\hyb, I)$ as a \textsf{hybrid instance} of $\alg$ if $\hyb$ is a hybrid algorithm of $\alg$ and $I$ is valid with respect to $\hyb$.
\end{definition}

The following lemma shows that by evaluating the cost difference between an MPFS algorithm $\alg$ and its hybrid algorithm $\hyb$, we can determine the competitive ratio of $\alg$.

\begin{lemma}
\label{lem-hyb-diff-comp}
The following claims hold for any $\alg\in\mc{MPFS}$.
\begin{enumerate}
    \item Suppose that $\alg(I)-\hyb(I)\le \alpha\cdot d(r_{\start},\stsvr)$ for any hybrid instance $H=(\hyb,I)=((\alg,\start,\stsvr),(S,d,\sigma))$ of $\omm(k)$. Then, $\alg$ is $(\alpha+1)$-competitive for $\omm(k)$.
    \item Furthermore, if $\alg(I)-\hyb(I) \le \alpha\cdot d_T^{\max}(r_{\start},\stsvr)$ for any hybrid instance $H=(\hyb,I)=((\alg,\start,\stsvr),(T,\sigma))$ of $\ommt(T)$,
then $\alg$ is T-strongly $(\alpha + |E(T)|)$-competitive for $\ommt(T)$.
\end{enumerate}
\end{lemma}

\begin{proof}
We first prove (1).
Let $I=(S,d,\sigma)$ be any instance for $\omm(k)$ and $\sigma=r_1\ldots r_k$. For $u=1,\ldots,k$, let $s_u\coloneqq\svrinst{u}{I}{\opt}$ and consider an algorithm $\alg_u$ defined as follows:
for $t=1,\ldots,u$, assign $r_t$ to $s_t$ and for $t=u+1,\ldots,k$, assign $r_t$ according to $\alg$.
Note that $\alg_k$ is just $\opt$ and let $\alg_0\coloneqq\alg$.
Then, we have
\[
\alg(I)-\opt(I)=\sum_{u=1}^k \left( \alg_{u-1}(I)-\alg_{u}(I) \right).
\]
Therefore, if $\alg_{u-1}(I)-\alg_{u}(I)\leq \alpha\cdot d(r_u, s_u)$ for $u=1,\ldots,k$, then we obtain
\begin{align*}
\alg(I)-\opt(I)&=\sum_{u=1}^k \left( \alg_{u-1}(I)-\alg_{u}(I) \right)\leq\alpha\cdot\sum_{u=1}^k d(r_u, s_u)=\alpha\cdot\opt(I),
\end{align*}
which implies that $\alg$ is $(\alpha+1)$-competitive.
Hence, it suffices to show that for $u=1,\ldots,k$,
\begin{equation}
\label{eq-hyb-comp}
\alg_{u-1}(I)-\alg_{u}(I)\leq \alpha\cdot d(r_u, s_u).
\end{equation}
If $\alg_{u-1}$ assigns $r_u$ to $s_u$, then obviously $\alg_{u-1}(I)-\alg_u(I)=0\leq \alpha\cdot d(r_u, s_u)$ holds.
Therefore, we prove (\ref{eq-hyb-comp}) only for the case where $\svrinst{u}{I}{\alg_{u-1}}\neq s_u$.

For $u=1,\ldots,k$, let $\sigma_u\coloneqq s_1\ldots s_{u-1}r_u\ldots r_k$ and
\[
H_u\coloneqq(\hyb_u,I_u)=((\alg,u,s_u),(S,d,\sigma_u)).
\] 
Since the total cost in the first $u-1$ steps is canceled out
in $\alg_{u-1}(I)-\alg_{u}(I)$, it follows that
\begin{align*}
\alg_{u-1}(I)-\alg_{u}(I)&=\alg(I_u)-\hyb_u(I_u)\le\alpha\cdot d(r_u,s_u),
\end{align*}
where the inequality is due to the assumption of $\alg(I)-\hyb(I)\le \alpha\cdot d(r_{\start},\stsvr)$.
This completes the proof of the first part.

Next, we prove (2).
Let $I=(T,\sigma)$ be any instance for $\ommt$ and $\sigma=r_1\ldots r_k$.
Similarly to the proof of the first part,
let $s_u\coloneqq \svrinst{u}{I}{\opt_T^{\max}}$ for $u=1,\ldots,k$ and $\alg_u$ be an algorithm that for $t=1,\ldots,u$, assigns $r_t$ according to $\opt_T^{\max}$ and for $t=u+1,\ldots,k$, assigns $r_t$ according to $\alg$. Let $\alg_0\coloneqq\alg$.
By the same arguments as in the first part, we have
$\alg_{u-1}(I)-\alg_{u}(I) \le \alpha \cdot d_T^{\max}(r_u,s_u)$
for $u=1,\ldots,k$.
By summing both sides from $u=1$ to $k$, we obtain
\begin{align*}
\alg(I)-\alg_{k}(I) \le \alpha\cdot\sum_{t=1}^k d_T^{\max}(r_t,s_t) = \alpha\cdot\opt_T^{\max}(I).
\end{align*}
Note that $\alg_{k}(I)=\sum_{t=1}^k d_T(r_t,s_t)$ is not the same value as $\opt_T^{\max}(I)=\sum_{t=1}^k d_T^{\max}(r_t,s_t)$.
By Remark~\ref{rem-two-distance-k-1}, we obtain $\alg_{k}(I) \le |E(T)|\cdot\opt_T^{\max}(I)$ and $\alg(I) \le (\alpha+|E(T)|)\opt_T^{\max}(I)$. This implies that $\alg$ is T-strongly $(\alpha+|E(T)|)$-competitive.
\end{proof}


Next, we introduce an important concept called \textit{cavities}~\cite{GL2012}, which is defined for a hybrid instance $H=(\hyb,I)=((\alg,\start,\stsvr),(S,d,\sigma))$ of $\omm$. First, consider the moment when $\alg$ assigns the $\start$-th request to $\svrinst{\start}{I}{\alg}$ and $\hyb$ assigns it to $\stsvr$. We regard $\alg$ as having an `extra' free server at $\stsvr$ (i.e., a server that is free for $\alg$ but full for $\hyb$). Similarly, we consider $\hyb$ to have an `extra' free server at $\svrinst{\start}{I}{\alg}$. The locations of these extra free servers may move as future requests arrive. Eventually, when both $\alg$ and $\hyb$ assign a request to their respective extra free servers, the sets of free servers of $\alg$ and $\hyb$ align, and from that point onward, $\alg$ and $\hyb$ assign a subsequent request to the same server. These extra free servers are referred to as \textit{cavities}.

The following lemma~\cite{HI2023} shows that for any hybrid instance $(\hyb,I)=((\alg,\start,\stsvr),(S,d,\sigma))$, the cavity of $\alg$ (i.e., a server that is free for $\alg$ but full for $\hyb$) and the cavity of $\hyb$ (a server that are free for $\hyb$ but full for $\alg$) are uniquely determined at each time, if they exist.
This holds by the property of MPFS algorithms.
%
%
\begin{lemma}[Harada and Itoh \mbox{\cite[Lemma 3]{HI2023}}]
\label{lem-hybrid-kihon}
Let $H=(\hyb,I)$ be a hybrid instance of $\alg\in\mc{MPFS}$,
where $\hyb=(\alg, \start, \stsvr)$.
Then, there uniquely exist a positive integer $\tend^H$ $(\geq \start)$ and
sequences of servers $\{h_t^H\}_{t=\start}^{\tend^H}$ and $\{a_t^H\}_{t=\start}^{\tend^H}$ such that
\begin{enumerate}

\item $F_t(\hyb,I)\setminus F_t(\alg,I)=\{h_t^H\}$ for each $\start\leq t\leq\tend^H$,

\item $F_t(\alg,I)\setminus F_t(\hyb,I)=\{a_t^H\}$ for each $\start\leq t\leq\tend^H$ and

\item $F_t(\alg,I)=F_t(\hyb,I)$ for each $\tend^H+1\leq t$.

\end{enumerate}
\end{lemma}

We call $\tend^H$ the \textit{\etime} of $H$ and refer to $h_t^H$ (resp. $a_t^H$)
as \textit{$\hyb$-cavity} (resp. \textit{$\alg$-cavity}) of $H$ at time $t$.
$\hyb$-cavities and $\alg$-cavities are collectively called \textit{cavities}.
In particular, $h_{\start}^H$ (resp. $a_{\start}^H$) is called
the \textit{first $\hyb$-cavity} (resp. \textit{the first $\alg$-cavity}) of $H$.
For simplicity, $\tend^H$, $h_t^H$ and $a_t^H$ are abbreviated as
$\tend$, $h_t$ and $a_t$ respectively
when $H$ is clear from the context.
We use $\cavity_{\hyb}(H)$ and $\cavity_{\alg}(H)$ to denote the sets of $\hyb$-cavities and $\alg$-cavities of $H$ respectively, i.e.,
\[
\cavity_{\hyb}(H)\coloneqq\{h_t\}_{t=\start}^{\tend},\:\: \cavity_{\alg}(H)\coloneqq\{a_t\}_{t=\start}^{\tend}.
\]
In addition, let $\cavity(H)\coloneqq\cavity_{\hyb}(H)\cup\cavity_{\alg}(H)$.
By the definition of cavities, it is easy to see that
the first $\alg$-cavity is the {\textdecsvr} of $\hyb$ and
the first $\hyb$-cavity is $\svrinst{{\start}}{I}{\alg}$.
%
%
%
%
%

We summarize the properties of cavities and assignments in the following proposition.
Intuitively, this proposition asserts the following:
\begin{itemize}
    \item At each time, either the $\hyb$-cavity or the $\alg$-cavity moves, or neither moves.
    \item If at time $t$, either one of the $\hyb$-cavity or the $\alg$-cavity moves, then we can determine the servers to which $\hyb$ and $\alg$ assigned requests at time $t$.
\end{itemize}
%
%
%
%
\begin{proposition}[Harada and Itoh \mbox{\cite[Proposition 1]{HI2023}}]
\label{prop-atht}
Let $H=(\hyb,I)$ be a hybrid instance of $\alg\in\mc{MPFS}$,
where $\hyb=(\alg,\start,\stsvr)$.
Then, the following properties hold:
\begin{enumerate}

\item $h_{t-1} = h_{t}$ or $a_{t-1} = a_{t}$ for each $\start+1\leq t \leq\tend $,

\item If $h_{t-1} \neq h_{t}$, then $r_{t}$ is assigned to $h_{t}$ by $\alg$
and to $h_{t-1}$ by $\hyb$ for each $\start+1 \leq t \leq \tend$,

\item If $a_{t-1} \neq a_{t}$, then $r_{t}$ is assigned to $a_{t-1}$ by $\alg$
and to $a_{t}$ by $\hyb$ for each $\start+1 \leq t \leq \tend$,

\item If $h_{t-1} = h_{t}$ and $a_{t-1} = a_{t}$, then $\alg$ and $\hyb$ assign $r_t$ to the same server and

\item $r_{\tend+1}$ is assigned to $a_{\tend}$ by $\alg$ and to $h_{\tend}$ by $\hyb$.

\end{enumerate}
\end{proposition}

As mentioned before, we aim to upper bound the difference in cost between an MPFS algorithm $\alg$ and its hybrid algorithm $\hyb$.
To this end, the hybrid cycle defined in the following definition plays a crucial role since the length of the hybrid cycle provides the upper bound on the difference in cost between $\alg$ and $\hyb$ (Lemma~\ref{lem-hyb-cyc-diff}).
For simplicity, we use the following notation for a distance function $d:X\times X\to \mathbb{R}_{\geq0}$ and $x_1,\ldots,x_n\in X$:
\[
\ring{d}(x_1,\ldots,x_n)\coloneqq d(x_1,x_n)+\sum_{i=1}^{n-1} d(x_i,x_{i+1}).
\]
%
%
%
%
\begin{definition}[Hybrid Cycle]
\label{def-hybcyc}
Let $H=(\hyb,I)$ be a hybrid instance of $\alg\in\mc{MPFS}$ for $\omm$, where
$\hyb=(\alg,\start,\stsvr)$ and $I=(S,d,\sigma)$.
We refer to the cycle
\[
h_{\start}\tct h_{\tend}\to a_{\tend}\tct a_{\start}\to h_{\start}
\]
as the \textsf{hybrid cycle} of $H$ and
define the \textsf{length} of the hybrid cycle to be
\begin{align*}
\ring{d}(H)&\coloneqq
\ring{d}(h_{\start},\ldots,h_{\tend},a_{\tend},\ldots,a_{\start})\\
&=d(h_{\start},a_{\start})+d(h_{\tend},a_{\tend})
+\sum_{t=\start+1}^{\tend} \left( d(h_{t-1},h_t)+d(a_{t-1},a_t) \right).
\end{align*}
\end{definition}
%
%


Applying Gupta and Lewi's method~\cite{GL2012} to MPFS algorithms yields the following lemma and its corollary.
The corollary implies that we can analyze the competitive ratio of MPFS algorithms by evaluating the length of the hybrid cycle.

\begin{lemma}
\label{lem-hyb-cyc-diff}
Let $\alg\in\mc{MPFS}$ and $H=(\hyb,I)=((\alg,\start,\stsvr),(S,d,\sigma))$ be any hybrid instance of $\omm(k)$.
Then, the difference in cost between $\alg$ and $\hyb$ is at most the length of the hybrid cycle, i.e.,
\[
\alg(I)-\hyb(I)\le \ring{d}(H).
\]
\end{lemma}

\begin{proof}
Fix any hybrid instance $H=(\hyb,I)=((\alg,\start,\stsvr),(S,d,\sigma))$ of $\omm(k)$ and let $\tend$ be the coupling time of $H$. For $t=\start,\ldots,k$,
let $\delta_t\coloneqq d(r_t,\svrinst{t}{I}{\alg})-d(r_t,\svrinst{t}{I}{\hyb})$. Note that
$\alg(I)-\hyb(I)=\sum_{t=\start}^k \delta_t$.
By the definition of $H$ and the triangle inequality, it follows that
\[
  \delta_{\start}=d(r_{\start},h_{\start})-d(r_{\start}, \stsvr)\leq d(h_{\start}, \stsvr) 
\]
and $\delta_t=0$ for each $t>\tend+1$.
By (5) of Proposition~\ref{prop-atht}, we have
\[
    \delta_{\tend+1}=d(r_{\tend+1},a_{\tend})-d(r_{\tend}, h_{\tend})\leq d(h_{\tend}, a_{\tend}).
\]
We then show that $\delta_t\leq d(h_{t-1},h_t)+d(a_{t-1},a_t)$ for $t=u+1,\ldots,\tend$.
If $h_{t-1}=h_t$ and $a_{t-1}=a_t$, then $\delta_t=0=d(h_{t-1},h_t)+d(a_{t-1},a_t)$ by (4) of Proposition~\ref{prop-atht}. 
If $h_{t-1}\neq h_t$, then $a_{t-1}=a_t$,
$\svrinst{t}{I}{\alg} =h_t$ and
$\svrinst{t}{I}{\hyb} =h_{t-1}$
hold by (1) and (2) of Proposition~\ref{prop-atht}.
By the triangle inequality, we get
\[
    \delta_t=d(r_t,h_t)-d(r_t, h_{t-1})\leq d(h_{t-1},h_t)+d(a_{t-1},a_t).
\]
If $a_{t-1}\neq a_t$, then $h_{t-1}=h_t$,
$\svrinst{t}{I}{\alg}=a_{t-1}$ and
$\svrinst{t}{I}{\hyb}=a_t$
hold by (1) and (3) of Proposition~\ref{prop-atht}.
Then,
\[
    \delta_t=d(r_t,a_{t-1})-d(r_t, a_{t})\leq d(h_{t-1},h_t)+d(a_{t-1},a_t).
\]
Putting the above all together, it follows that
\begin{align*}
\alg(I)-\hyb(I)&
=\sum_{t=\start}^k \delta_t
\leq d(h_{\start}, a_{\start})+d(h_{\tend}, a_{\tend})+
\sum_{t=\start+1}^{\tend} \left\{d(h_{t-1},h_t)+d(a_{t-1},a_t)\right\} \\
&=\ring{d}(h_u,\ldots,h_{\tend},a_{\tend},\ldots,a_u)=\ring{d}(H).
\end{align*}
This completes the proof.
\end{proof}

By Lemmas~\ref{lem-hyb-diff-comp} and~\ref{lem-hyb-cyc-diff}, we immediately have the following corollary.
%
%
%
%
\begin{corollary}
\label{cor-hyb-comp}
The following claims hold for any $\alg\in\mc{MPFS}$.
\begin{enumerate}
    \item If $\ring{d}(H)\leq \alpha\cdot d(r_{\start},\stsvr)$ for any hybrid instance $H=(\hyb,I)=((\alg,\start,\stsvr),(S,d,\sigma))$ of $\omm(k)$, then $\alg$ is $(\alpha+1)$-competitive for $\omm(k)$.
    \item If $\ring{d}_T(H) \le \alpha\cdot d_T^{\max}(r_{\start},\stsvr)$ for any hybrid instance $H=(\hyb,I)=((\alg,\start,\stsvr),(T,\sigma))$ of $\ommt(T)$, then $\alg$ is T-strongly $(\alpha + |E(T)|)$-competitive for $\ommt(T)$.
\end{enumerate}
\end{corollary}

By this corollary, we aim to obtain an inequality of the form $\ring{d}(H)\leq \alpha\cdot d(r_{\start},\stsvr)$, and focus on hybrid instances that have a certain first $\hyb$-cavity and $\alg$-cavity.
Then, we introduce the following notation:
%
%
%
%
%
%
%
%
%
For $\omm$ with a set $S$ of servers and a distance function $d$,
we use $\mf{H}_{S,d}^{\alg}(h,\stsvr)$ to denote a set of hybrid instances $H=(\hyb,I)$ of $\alg$
such that the first $\hyb$-cavity is $h$ and the first $\alg$-cavity (or the {\textdecsvr} of $\hyb$) is $\stsvr$, i.e.,
%
%
\[
\mf{H}_{S,d}^{\alg}(h,\stsvr)
\coloneqq\{H=((\alg,\start,\stsvr),(S,d,\sigma)):h_{\start}=h\}.
\]
For $\ommt(T)$, we use $\mf{H}_T^{\alg}(h,\stsvr)$ instead of
$\mf{H}_{V(T),d_T}^{\alg}(h,\stsvr)$, where $V(T)$ is the vertex set of $T$ and $d_T$ is the path distance on $T$.
The superscript $\alg$ may be omitted when $\alg$ is clear from the context.

The following definition describes a good property about a hybrid instance called \textit{well-behaved}.
In a well-behaved hybrid instance, until the {\stime}, all requests occur at a location of a free server. From the {\stime} onwards, the cavity moves at each time, and the sets of free servers of $\hyb$ and $\alg$ do not coincide until the last request arrives.
%
%
%
%
\begin{definition}[Well-Behaved Hybrid Instance]
\label{def-well-behaved}
Let $H=(\hyb,I)$ be a hybrid instance for $\omm(k)$,
where $\hyb=(\alg,\start,\stsvr)$, $I=(S,d,\sigma)$ and $\sigma=r_1\ldots r_k$.
We say that $H$ is \textsf{well-behaved} if $H$ satisfies the following conditions:

\begin{enumerate}

\item The {\etime} of $H$ is $k-1$, i.e., $\tend=k-1$,
\label{wb-k-1}

\item $d(r_t, \svrinst{t}{I}{\alg})=0$ for any $1\leq t \leq \start-1$,
\label{wb-drs0}

\item $F_t(\alg,I)=\cavity_t(H)\setminus\{h_t\}$ for any $\start\leq t\leq k-1$ and
\label{wb-freeA}

\item $F_t(\hyb,I)=\cavity_t(H)\setminus\{a_t\}$ for any $\start\leq t\leq k-1$,
\label{wb-freeH}

\end{enumerate}
where $\cavity_t(H)$ denotes the set of $\hyb$-cavities and $\alg$-cavities from time $t$, i.e., $\cavity_t(H)\coloneqq\bigcup_{u=t}^{k-1}\{h_u,a_u\}$.
\end{definition}
\begin{remark}
\label{rem-not-well-behaved}
For each (not necessarily well-behaved) hybrid instance $H$ for $\omm(k)$, it follows that
$F_t(\alg,I)\supseteq\cavity_t(H)\setminus\{h_t\}$ and
$F_t(\hyb,I)\supseteq\cavity_t(H)\setminus\{a_t\}$.
\end{remark}
%
%
%
%
\begin{remark}
\label{rem-well-behaved}
By (\ref{wb-freeA}) and (\ref{wb-freeH}) in Definition~\ref{def-well-behaved},
we can observe that for any $\start+1\leq t\leq k-1$,
$h_{t-1}\neq h_t$ or $a_{t-1}\neq a_t$.
\end{remark}

The following lemma claims that
for any hybrid instance $H$,
there exists a well-behaved hybrid instance $H'$
such that
the `movement of cavities' in $H'$ is identical to that in $H$.
Thanks to this lemma,
we can conclude that
for almost all propositions and lemmas concerning hybrid instances in this paper,
it suffices to only consider well-behaved hybrid instances when proving them.
%
%
%
%
\begin{lemma}
\label{lem-well-behaved}
Let
$H=(\hyb,I)=((\alg,\start,\stsvr),(S,d,\sigma))\in\mf{H}_{S,d}(h,\stsvr) $
be any hybrid instance of $\alg$ for $\omm$
with 
$\cavity_{\hyb}(H)=\{ h_t \}_{t=\start}^{\tend} $
and
$\cavity_{\alg}(H)=\{ a_t \}_{t=\start}^{\tend} $.
Let $t(1)<\cdots < t(n-1)$ denote values of $t$ such that
$h_{t-1}\neq h_t$ or $a_{t-1}\neq a_t$,
and let
$t(0)\coloneqq\start$.
Then, there exists a well-behaved hybrid instance
$H'=(\hyb',I')=((\alg,\start',\stsvr),(S,d,\sigma'))$
such that
$ h'_{\start'+i}=h_{t(i)} $
and
$ a'_{\start'+i}=a_{t(i)} $
for each $i=0,\ldots,n-1$,
where $h'_t$ and $a'_t$ denote the $\hyb'$-cavity and $\alg$-cavity of $H'$ at time $t$
respectively.
Moreover, for such $H'$,
$H'\in\mf{H}_{S,d}(h,\stsvr)$,
$\cavity(H)=\cavity(H')$
and
$\ring{d}(H)=\ring{d}(H')$
hold.
\end{lemma}

%
%
%
%
%
%
%

\begin{proof}
Let
$H=(\hyb,I)=((\alg,\start,\stsvr),(S,d,\sigma)) \in \mf{H}_{S,d}(h,\stsvr)$
be any hybrid instance of $\alg$ for $\omm(k)$, where $\sigma=r_1\ldots r_k$.
We construct a well-behaved hybrid instance $H'=(\hyb',I')=((\alg,\start',\stsvr),(S,d,\sigma'))$ such that $H'\in\mf{H}_{S,d}(h,\stsvr)$, $\cavity(H)=\cavity(H')$ and $\ring{d}(H)=\ring{d}(H')$.
We use $\{h_t\}_{t=\start}^{\tend}$ and $\{a_t\}_{t=\start}^{\tend}$ to denote the $\hyb$-cavities and $\alg$-cavities of $H$ respectively, and similarly, 
$\{h'_t\}_{t=\start'}^{k-1}$ and $\{a'_t\}_{t=\start'}^{k-1}$ 
to denote the $\hyb'$-cavities and $\alg$-cavities of $H'$ respectively.
For convenience, let $t(n)\coloneqq\tend+1$.

Since two cavities exist at time $t(0)=\start$ and one new cavity appears at each time $t(i)$ for $i=1,\ldots,n-1$, we have $|\cavity(H)|=n+1$.
In addition, by the definition of $t(1),\ldots,t(n-1)$ and $t(n)$, we have
\begin{align*}
    \ring{d}(H)&=\ring{d}(h_{\start},\ldots,h_{\tend},a_{\tend},\ldots,a_{\start}) \\
    &=\ring{d}(h_{t(0)},\ldots,h_{t(n-1)},a_{t(n-1)},\ldots,a_{t(0)}).
\end{align*}

We set $\start'\coloneqq k-n$ and define $\sigma'$ as follows:
First, $k-n-1$ requests arrive on each server $s\in S\setminus\cavity(H)$
and then requests from $r_{t(0)}$ to $r_{t(n)}$ arrive.
In what follows, we show that
for $i=0,\ldots,n-1$,
\begin{align}
\label{eq-fcav1}
F_{\start'+i}(\hyb',I')&=\cavity_{t(i)}(H)\setminus\{a_{t(i)}\} \text{ and }\\
F_{\start'+i}(\alg,I')&=\cavity_{t(i)}(H)\setminus\{h_{t(i)}\}.
\label{eq-fcav2}
\end{align}

\noindent
Note that
$
\cavity_{t(i)}(H)
=\cavity_{t(i+1)-1}(H)
=\bigcup_{j=i}^{n-1} \{ h_{t(j)},a_{t(j)} \}
$
by the definition of $t(\cdot)$.
If we establish (\ref{eq-fcav1}) and (\ref{eq-fcav2}),
then $h'_{\start'+i}=h_{t(i)}$ and $a'_{\start'+i}=a_{t(i)}$ hold for $i=0,\ldots,n-1$.
This implies that the following claims hold and the proof is completed.
\begin{itemize}
\item $H'$ is well-behaved,

\item $H'\in\mf{H}_{S,d}(h_{t(0)},a_{t(0)})=\mf{H}_{S,d}(h,\stsvr)$,

\item $\cavity(H)=\bigcup_{i=0}^{n-1} \{ h_{t(i)},a_{t(i)} \}=\bigcup_{i=0}^{n-1} \{ h'_{\start'+i},a'_{\start'+i} \}=\cavity(H')$ and

\item $\ring{d}(H)=\ring{d}(h_{t(0)},\ldots,h_{t(n-1)},a_{t(n-1)},\ldots,a_{t(0)})
=\ring{d}(h'_{\start'},\ldots,h'_{k-1},a'_{k-1},\ldots,a'_{\start'})=\ring{d}(H')$.
\end{itemize}

To prove (\ref{eq-fcav1}) and (\ref{eq-fcav2}), we use induction on $i$.

\half
\noindent
\textbf{Base case $i=0$:}
\quad
By the definition of $\sigma'$, we have
\begin{align}
\label{eq-lemma-well-behaved-basecase}
F_{\start'-1}(\alg,I')=F_{\start'-1}(\hyb',I')=\cavity(H)=\cavity_{t(0)}(H).
\end{align}
Since $a_{t(0)}=a_{\start}=\stsvr$ and $\hyb'$ assigns the $\start'$-th request to $\stsvr$ when processing $I'$, we have
\[
F_{\start'}(\hyb',I')
=F_{\start'-1}(\hyb',I')\setminus\{\stsvr\}
=F_{\start'-1}(\hyb',I')\setminus\{a_{t(0)}\}
=\cavity_{t(0)}(H)\setminus\{a_{t(0)}\}
\]
and (\ref{eq-fcav1}) holds for $i=0$.
Next, we show (\ref{eq-fcav2}).
Since $\alg\in\mc{MPFS}$, we have

\begin{itemize}
\item $h_{t(0)}=h_{\start}=
\svrmpfs{r_{\start}}{F_{\start-1}(\alg,I)}{\alg}
$,

\item $F_{\start'-1}(\alg,I')=\cavity(H)\subseteq F_{\start-1}(\alg,I)$ and

\item $h_{t(0)}=h_{\start}\in\cavity(H)=F_{\start'-1}(\alg,I')$.
\end{itemize}

\noindent
Recall that $\svrmpfs{r}{F}{\alg}$ denotes the server to which $\alg$ assigns when a new request is $r$ and the set of free servers is $F$.
By Remark~\ref{rem-mpfs}, we get
$
h'_{\start'}
=\svrmpfs{r_{t(0)}}{F_{\start'-1}(\alg,I)}{\alg}
=h_{t(0)}
$
and
\[
F_{\start'}(\alg,I')=F_{\start'-1}(\alg,I')\setminus\{h'_{\start'}\} =F_{\start'-1}(\hyb',I')\setminus\{h_{t(0)}\}=\cavity_{t(0)}(H)\setminus\{h_{t(0)}\},
\]
where the last equality is due to (\ref{eq-lemma-well-behaved-basecase}).
Therefore, (\ref{eq-fcav2}) holds for $i=0$.

\half
\noindent
\textbf{Inductive step:}
\quad Assume that (\ref{eq-fcav1}) and ($\ref{eq-fcav2}$) hold for $i-1$.
By the definition of $t(\cdot)$, we have $h_{t(i-1)}=h_{t(i)-1}$, $a_{t(i-1)}=a_{t(i)-1}$, and either $h_{t(i)-1}\neq h_{t(i)}$ or $a_{t(i)-1}\neq a_{t(i)}$.
Since we can show the first case $h_{t(i)-1}\neq h_{t(i)}$ and the second case $a_{t(i)-1}\neq a_{t(i)}$ analogously, we prove only the first case.

By (2) of Proposition~\ref{prop-atht}, we have $h_{t(i)-1}=\svrmpfs{r_{t(i)}}{F_{t(i)-1}(\hyb, I)}{\alg}$ and the following:
\begin{itemize}
    \item $\cavity_{t(i)-1}(H) \setminus \{a_{t(i)-1}\}\subseteq F_{t(i)-1}(\hyb, I)$ and
    \item $h_{t(i-1)}=h_{t(i)-1}\in \cavity_{t(i)-1}(H) \setminus \{a_{t(i)-1}\}=F_{\start'+i-1}(\hyb',I')$, 
\end{itemize}
where the second inequality is due to the induction hypothesis.
By Remark~\ref{rem-mpfs}, $\hyb'$ assigns $r_{t(i)}$ to $h_{t(i-1)}$ for $I'$.
Then, we have
\begin{align*}
F_{\start'+i}(\hyb',I')
&=F_{\start'+i-1}(\hyb',I')\setminus\{h_{t(i-1)}\} \\
&=\left(\cavity_{t(i-1)}(H)\setminus\{a_{t(i-1)}\}\right)\setminus\{h_{t(i-1)}\} \\
&=\left(\cavity_{t(i)}(H)\setminus\{a_{t(i-1)}\}\right)\setminus\{h_{t(i-1)}\} \\
&=\cavity_{t(i)}(H)\setminus\{a_{t(i)}\},
\end{align*}
where the second equality is due to the induction hypothesis, the third equality is due to $h_{t(i-1)}, a_{t(i-1)}\in \cavity_{t(i-1)}$ and the last equality is due to $h_{t(i-1)}\notin\cavity_{t(i)}(H)$ and $a_{t(i-1)}=a_{t(i)}$.

By Remark~\ref{rem-mpfs} and
$
h_{t(i)}
\in \cavity_{t(i)-1}(H) \setminus \{h_{t(i)-1}\}
\subseteq F_{t(i)-1}(\alg, I)
$,
we can similarly show that $\alg$ assigns $r_{t(i)}$ to $h_{t(i)}$ for $I'$.
Then, it follows that
\begin{align*}
F_{\start'+i}(\alg,I')
&=F_{\start'+i-1}(\alg,I')\setminus\{h_{t(i)}\} \\
&=\left(\cavity_{t(i-1)}(H)\setminus\{h_{t(i-1)}\}\right)\setminus\{h_{t(i)}\} \\
&=\left(\cavity_{t(i)}(H)\cup\{a_{t(i-1)}\}\right)\setminus\{h_{t(i)}\} \\
&=\cavity_{t(i)}(H)\setminus\{h_{t(i)}\},
\end{align*}
where the second equality is due to the induction hypothesis and the last equality is due to $a_{t(i-1)}=a_{t(i)}$.
\end{proof}

%
%
%
%
%
%
%
Next, we define a conjugate instance of a hybrid instance $H$ and prove its existence.
In a conjugate instance $H'$ of $H$, all $\hyb$-cavities of $H$ is $\alg$-cavities of $H'$ and all $\alg$-cavities of $H$ is $\hyb$-cavities of $H'$, effectively reversing the roles of $\alg$ and $\hyb$.
By introducing this definition, we can avoid repetitive arguments when proving the important lemma (Lemma~\ref{lem-hybrid-main1}) at the end of this section.
%
%
%
%
\begin{definition}[Conjugate Instance]
\label{def-conjugate}
    Let $H=(\hyb,I)=((\alg,\start,\stsvr),(S,d,\sigma))$ be a hybrid instance.
    We say that
    $H'=(\hyb',I')=((\alg,\start,h_{\start}^H),(S,d,\sigma'))$
    is a \textsf{conjugate} instance of $H$ if the following conditions hold:
    \begin{enumerate}
    \setlength{\leftskip}{1.0cm}
        \item $\tend^{H}=\tend^{H'}$,
        \item $h_t^H=a_t^{H'}$ and $a_t^H=h_t^{H'}$ for any $\start\leq t\leq\tend^H=\tend^{H'}$.
    \end{enumerate}    
\end{definition}

\begin{lemma}
\label{lem-conjugate}
For any hybrid instance $H=(\hyb,I)=((\alg,\start,\stsvr),(S,d,\sigma))$,
there exists a conjugate instance $H'=(\hyb',I')=((\alg,\start,h_{\start}^H),(S,d,\sigma'))$ of $H$.
\end{lemma}

\begin{proof}
Let $\sigma'$ be the request sequence obtained by changing $r_{\start}$ to $\stsvr$.
Then, evidently
$F_{\start-1}(\alg,I)=F_{\start-1}(\alg,I')=F_{\start-1}(\hyb,I)=F_{\start-1}(\hyb',I')$ holds since the first $\start-1$ requests of $\sigma$ and $\sigma'$ are the same.
By the definition of hybrid algorithms and $\sigma'$, we have
\begin{align*}
F_{\start}(\alg,I)&=F_{\start-1}(\alg,I)\setminus\{h_{\start}^H\}
=F_{\start-1}(\hyb',I')\setminus\{h_{\start}^H\}=F_{\start}(\hyb',I') \text{ and}\\
F_{\start}(\hyb,I)&=F_{\start-1}(\hyb,I)\setminus\{\stsvr\}
=F_{\start-1}(\alg,I')\setminus\{\stsvr\}=F_{\start}(\alg,I').
\end{align*}
Since $\alg$ is in $\mc{MPFS}$ and $\sigma'$ is the same as $\sigma$ except for the $\start$-th request,
it follows that for $t\geq\start$,
\begin{align*}
F_{t}(\alg,I)=F_{t}(\hyb',I') \text{ and }
F_{t}(\hyb,I)=F_{t}(\alg,I').
\end{align*}
This implies that $h_t^H=a_t^{H'}$ and $a_t^H=h_t^{H'}$ for any $\start\leq t\leq\tend^H=\tend^{H'}$.
\end{proof}
The following lemma is the most important in this section. This lemma claims that there exists a hybrid instance that has a `partial cycle' of a certain hybrid cycle. This lemma is useful when upper bounding the length of hybrid cycles for a recursively defined MPFS algorithm $\alg$. It allows us to use upper bounds on the length of hybrid cycles for algorithms used as subroutines in $\alg$ as an induction hypothesis. In this paper, particularly, it plays a significant role in the proof of Lemma~\ref{lem-main1} (Appendix~\ref{app-proof-main-lemma}).
%
%
%
%
\begin{lemma}
\label{lem-hybrid-main1}
For any hybrid instance $H=(\hyb,I)=((\alg,\start,\stsvr),(S,d,\sigma))$,
let $\tend$, $\{h_t\}_{t=\start}^{\tend}$ and $\{a_t\}_{t=\start}^{\tend}$ be the {\etime}, $\hyb$-cavities and $\alg$-cavities of $H$ respectively. 
Then, the following properties hold:
\begin{enumerate}

\item For any $t_1,t_2$ with $\start\leq t_1<t_2\leq\tend$, there exists a hybrid instance $H'=(\hyb',I')=((\alg,\start',h_{t_2}),(S,d,\sigma'))\in\mf{H}_{S,d}(h_{t_1},h_{t_2})$ such that $\cavity(H')=\{h_{t_1},\ldots,h_{t_2}\}$ and
\[
\ring{d}(H')=\ring{d}(h_{t_1},\ldots,h_{t_2}).
\]
\item For any $t_1,t_2$ with $\start\leq t_1<t_2\leq\tend$, there exists a hybrid instance $H'=(\hyb',I')=((\alg,\start',a_{t_2}),(S,d,\sigma'))\in\mf{H}_{S,d}(a_{t_1},a_{t_2})$ such that $\cavity(H')=\{a_{t_1},\ldots,a_{t_2}\}$ and
\[
\ring{d}(H')=\ring{d}(a_{t_1},\ldots,a_{t_2}).
\]

\item For any $t_1,t_2$, there exists a hybrid instance $H'=(\hyb',I')=((\alg,\start',a_{t_2}),(S,d,\sigma'))\in\mf{H}_{S,d}(h_{t_1},a_{t_2})$ such that $\cavity(H')=\{h_{t_1},\ldots,h_{\tend},a_{\tend},\ldots,a_{t_2}\}$ and
\[
\ring{d}(H')=\ring{d}(h_{t_1},\ldots,h_{\tend},a_{\tend},\ldots,a_{t_2}).
\]

\end{enumerate}
\end{lemma}

\noindent
Before we provide a rigorous proof, let us first provide intuitive explanation for (1).
In particular, we explain what $H'=(\hyb',I')=((\alg,\start',h_{t_2}),(S,d,\sigma'))\in\mf{H}_{S,d}(h_{t_1},h_{t_2})$ is. 
For simplicity, assume that
$h_{t_1},\ldots,h_{t_2} $ differ from each other.
Let $\start'=k-|\{h_{t_1},\ldots,h_{t_2} \}|+1$ and $\sigma'$ defined as follows:
$\start'-1$ requests arrive on each server $s\in S\setminus\{h_{t_1},\ldots,h_{t_2} \}$ at first,
and then requests from $r_{t_1}$ to $r_{t_2}$ arrive.
For this hybrid instance $H'$, we can see that
(i) the first $\hyb$-cavity is $h_{t_1}$ and the second is $h_{t_1+1}$, and so on, and
(ii) the first $\alg$-cavity is $h_{t_2}$ and the $\alg$-cavity does not change after that.

\begin{proof}[Proof of Lemma~\ref{lem-hybrid-main1}]
First, we prove that it suffices to show the lemma holds only for well-behaved hybrid instances.
Since we can similarly show this for (1), (2) and (3),
we omit the proofs for (2) and (3).
Suppose that (1) of the lemma holds for any well-behaved hybrid instance and
let
$H=(\hyb,I)=((\alg,\start,\stsvr),(S,d,\sigma))$
be any hybrid instance.
By Lemma~\ref{lem-well-behaved},
there exists a well-behaved hybrid instance $H'=(\hyb',I')$ such that
$h'_{\start'+i}=h_{t(i)}$
and
$a'_{\start'+i}=a_{t(i)}$,
where
the notation follows the one used in Lemma~\ref{lem-well-behaved}.
By the definition of $t(\cdot)$,
it follows that for any $t_1$, $t_2$ with $\start \leq t_1 < t_2 \leq \tend$, there exist
\textit{unique} positive integers $i_1<i_2$ such that
$h_{t(i_1)}=h_{t_1}$
and
$h_{t(i_2)}=h_{t_2} $.
If we apply the lemma for $H'$,
then there exists
$H''\in\mf{H}_{S,d}(h'_{\start'+i_1},h'_{\start'+i_2})$ such that
$\cavity(H'')=\{h'_{\start'+i_1},\ldots,h'_{\start'+i_2}\}$
and
$\ring{d}(H'')=\ring{d}(h'_{\start'+i_1},\ldots,h'_{\start'+i_2}) $.
For such $H''$, we have
\begin{align*}
    (h'_{\start'+i_1},h'_{\start'+i_2})
    &= (h_{t(i_1)},h_{t(i_2)})
    = (h_{t_1},h_{t_2}),\\
    \cavity(H'')
    &=\{h_{t(i_1)},\ldots,h_{t(i_2)}\}
    =\{h_{t_1},\ldots,h_{t_2}\} \text{ and }\\
    \ring{d}(H'')
    &=\ring{d}(h_{t(i_1)},\ldots,h_{t(i_2)})
    =\ring{d}(h_{t_1},\ldots,h_{t_2}).
\end{align*}
This implies that (1) of the lemma also holds for $H$.

Next,
for any well-behaved hybrid instance $H=(\hyb,I)=((\alg,\start,\stsvr),(S,d,\sigma))$ with
$\cavity_{\hyb}(H)=\{h_t\}_{t=\start}^{\tend}$ and
$\cavity_{\alg}(H)=\{a_t\}_{t=\start}^{\tend}$,
we show (1)
by constructing
$H'=(\hyb',I')=((\alg,\start',h_{t_2}),(S,d,\sigma'))\in\mf{H}_{S,d}(h_{t_1},h_{t_2})$.
Let $t(1)<\cdots <t(n)$ be values of $t$ such that $h_{t-1}\neq h_t$ and $t_1+1\le t\le t_2$, and
let $t(0)$ be a value of $t$ such that $h_{t-1}\neq h_t=h_{t_1}$.
Then, it follows that
\[
    \{h_{t_1},\ldots,h_{t_2} \}
    = \{h_{t(0)},\ldots,h_{t(n)} \}
    \text{ and }
    \ring{d}(h_{t_1},\ldots,h_{t_2})
    = \ring{d}(h_{t(0)},\ldots,h_{t(n)}).
\]

\noindent
We set $\start'\coloneqq k-n$ and define $\sigma'$ as follows:
first, $k-n-1$ requests arrive on each server $s\in S\setminus\{h_{t_1},\ldots,h_{t_2} \}$ and then requests from $r_{t(0)}$ to $r_{t(n)}$ arrive.

If we show that, for $i=0,\ldots,n-1$,
\begin{align}
    \label{eq-lem-hyb-main1}
    F_{\start'+i}(\hyb',I')&=\{h_{t(i)},\ldots,h_{t(n-1)} \}
    \text{ and} \\
    \label{eq-lem-hyb-main2}
    F_{\start'+i}(\alg,I')&=\{h_{t(i+1)},\ldots,h_{t(n)} \},
\end{align}
then
$h'_{\start'+i}=h_{t(i)} $
and
$a'_{\start'+i}=h_{t(n)} $
hold.
This implies that
\begin{align*}
    \cavity(H')&= \{ h_{t(0)},\ldots,h_{t(n)} \}=\{ h_{t_1},\ldots,h_{t_2} \} \text{ and } \\
    \ring{d}(H')&= \ring{d}(h_{t(0)},\ldots,h_{t(n)}).
\end{align*}
Therefore, in what follows, we show (\ref{eq-lem-hyb-main1}) and (\ref{eq-lem-hyb-main2}) by induction on $i$.

\half
\noindent
\textbf{Base case $i=0$:}
\quad
(\ref{eq-lem-hyb-main1}) holds for $i=0$ by the following calculation:
\[
    F_{\start'}(\hyb',I')
    =\{ h_{t(0)},\ldots,h_{t(n)} \}\setminus\{h_{t_2} \}
    =\{ h_{t(0)},\ldots,h_{t(n-1)} \}.
\]

\noindent
Since $H$ is well-behaved, we have
$
    F_{t(0)-1}(\alg,I)
    =\cavity_{t(0)-1}(H)\setminus\{h_{t(0)-1} \}.
$
Then, we obtain
\begin{align*}
    h_{t(0)}
    =\svrmpfs{r_{t(0)}}{F_{t(0)-1}(\alg, I)}{\alg}
    \text{ and }
    h_{t(0)}
    \in F_{\start'-1}(\alg, I')
    \subseteq F_{t(0)-1}(\alg,I).
\end{align*}
This leads to
$
h_{t(0)}
=\svrmpfs{r_{t(0)}}{F_{\start'-1}(\alg, I')}{\alg}
=h'_{\start'} $.
Thus, we have
\[
    F_{\start'}(\alg,I')
    =\{ h_{t(0)},\ldots,h_{t(n)} \}\setminus\{h_{t(0)} \}
    =\{ h_{t(1)},\ldots,h_{t(n-1)} \},
\]
in other words, (\ref{eq-lem-hyb-main2}) holds for $i=0$.

\half
\noindent
\textbf{Inductive step:}
\quad
Assume that (\ref{eq-lem-hyb-main1}) holds for $i-1$.
Since $H$ is well-behaved, 
$
    F_{t(i)-1}(\alg,I)
    =\cavity_{t(i)-1}(H)\setminus\{h_{t(i)-1} \}
$
and
$
    F_{t(i)-1}(\hyb,I)
    =\cavity_{t(i)-1}(H)\setminus\{a_{t(i)-1} \}
$
hold.
By (2) of Proposition~\ref{prop-atht}, we have
$
    h_{t(i-1)}=h_{t(i)-1}
    =\svrmpfs{r_{t(i)}}{F_{t(i)-1}(\hyb, I)}{\alg}
$
and
\begin{align*}
    h_{t(i-1)}
    \in \{ h_{t(i-1)},\ldots, h_{t(n-1)} \}
    \subseteq \cavity_{t(i)-1}(H)\setminus\{a_{t(i)-1}\}
    = F_{t(i)-1}(\hyb, I) .
\end{align*}
Since
$
    \{ h_{t(i-1)},\ldots, h_{t(n-1)} \}
    =F_{\start'+i-1}(\hyb',I')
$
by the induction hypothesis, we can see that
$\hyb'$ assigns $r_{t(i)}$ to $h_{t(i-1)}$ for $I'$
(see Remark~\ref{rem-mpfs}).
Then, (\ref{eq-lem-hyb-main1}) holds for $i$.
Next, suppose that (\ref{eq-lem-hyb-main2}) holds for $i-1$.
Similarly to the above discussion, we can also see that
$\alg$ assigns $r_{t(i)}$ to $h_{t(i)}$ for $I'$, that is, (\ref{eq-lem-hyb-main2}) holds for $i$.
This completes the proof of (1) of the lemma.

Once (1) is proved, we can easily show (2) by using a conjugate instance of $H$.
By Lemma~\ref{lem-conjugate}, there exists a conjugate instance $H'=(\hyb',I')$ of $H$.
Let $h'_t$ (resp. $a'_t$) be the $\hyb'$-cavity (resp. $\alg$-cavity) of $H'$ at time $t$.
By the definition of conjugate instances, it follows that
$h_t=a'_t$ and $a_t=h'_t$
for any $t_1\leq t\leq t_2$.
Then, by applying (1) for $H'$, there exists a hybrid instance $H''$ such that
$\cavity(H'')=\{h'_{t_1},\ldots,h'_{t_2}\}=\{a_{t_1},\ldots,a_{t_2} \}$
and
$\ring{d}(H'')=\ring{d}(h'_{t_1},\ldots,h'_{t_2})=\ring{d}(a_{t_1},\ldots,a_{t_2}) $.
This implies that (2) holds.

Finally, we show (3) for the case $t_1\leq t_2$.
If we can show (3) for the case $t_1\leq t_2$,
then, by using a conjugate instance, (3) for the case $t_1 > t_2$ is easily proved.
Let $t(1)<\cdots < t(n-1)$ be values of $t$ such that
$h_{t-1}\neq h_t$ and $t_1+1\leq t\leq t_2$,
and let $t(0)$ be a value of $t$ such that $h_{t-1}\neq h_t=h_{t_1}$.
By the definition of $t(\cdot)$ and well-behaved hybrid instances, we have the following:
\begin{itemize}
    \item $h_{t(n-1)}=h_{t_2} $,
    \item $\{h_{t(0)},\ldots,h_{t(n-2)},h_{t(n-1)},h_{t_2+1},\ldots,h_{\tend},a_{\tend},\ldots,a_{t_2}\}
    =\{h_{t_1},\ldots,h_{\tend},a_{\tend},\ldots,a_{t_2}\}$ and
    \item $|\{h_{t_1},\ldots,h_{\tend},a_{\tend},\ldots,a_{t_2}\}|=k-(t_2-n)$,
\end{itemize}
where the third equality is due to $\tend=k-1$.
Let $\start'\coloneqq t_2-n+1$ and $\sigma'$ defined as follows:
\begin{enumerate}
    \item $\start'-1$ requests arrive on each server $s\in S\setminus\{h_{t_1},\ldots,h_{\tend},a_{\tend},\ldots,a_{t_2}\}$, and then
    \item requests $r_{t(0)}r_{t(1)}\ldots r_{t(n-1)}r_{t_2}r_{t_2+1}\ldots r_{k-1}r_{k}$ arrive from $r_{t(0)}$ to $r_k$.
\end{enumerate}
In the same way as the proof of (1), we can see that, for $i=0,\ldots,n-1$,
\begin{align*}
    F_{\start'+i}(\hyb',I')&=\{h_{t(i)},\ldots,h_{t(n-1)},h_{t_2+1},\ldots,h_{\tend},a_{\tend},\ldots,a_{t_2-1} \}
    \text{ and} \\
    F_{\start'+i}(\alg,I')&=\{h_{t(i+1)},\ldots,h_{t(n-1)},h_{t_2+1},\ldots,h_{\tend},a_{\tend},\ldots,a_{t_2} \},
\end{align*}
which means that $h'_{\start'+i}=h_{t(i)}$ and $a'_{\start'+i}=a_{t_2}$ hold for $i=0,\ldots,n-1$.
In addition, we also have
\begin{align*}
    \start'+n-1&=t_2, \\
    F_{t_2}(\hyb',I')&=\{h_{t_2},\ldots,h_{\tend},a_{\tend},\ldots,a_{t_2-1} \}=\cavity_{t_2}(H)\setminus\{a_{t_2}\}=F_{t_2}(\hyb,I)
    \text{ and} \\
    F_{t_2}(\alg,I')&=\{h_{t_2+1},\ldots,h_{\tend},a_{\tend},\ldots,a_{t_2} \}=\cavity_{t_2}(H)\setminus\{h_{t_2}\}=F_{t_2}(\alg,I),
\end{align*}
where the second equality is due to $h_{t(n-1)}=h_{t_2}$
and the fourth and last equalities are due to the definition of well-behaved hybrid instances.
Moreover, since $\sigma$ and $\sigma'$ have the same requests at time $t\geq t_2$,
$H'$ has the same cavities as $H$ does at $t \geq t_2$.
This implies that $\cavity(H')=\{h_{t_1},\ldots,h_{\tend},a_{\tend},\ldots,a_{t_2}\}$ and
$\ring{d}(H')=\ring{d}(h_{t_1},\ldots,h_{\tend},a_{\tend},\ldots,a_{t_2})$.
\end{proof}
%
%
%
%
%

%
%
%
%
%

%
%

%
%
%
%
%

\section{Hybrid Algorithm of Subtree-Decomposition}
\label{sec-hybrid-algorithm-sd}

In this section, we present two lemmas which play important roles in deriving the upper bound on the competitive ratio of the SD algorithm $\algs$.

%
%
%
%
%
%
\begin{definition}
\label{def-simulate}
Let $U\in \{T_i\}_{i=1}^l \cup \{T_{-i}\}_{i=1}^l \cup \{T^{(1)},T^{(2)}\}$ be a subtree of $T$ and $H\in\mf{H}_T(h,\stsvr)$ be a hybrid instance of $\algs$ for $\ommt(T)$.
We say that a hybrid instance $H'\in\mf{H}_U(h,\stsvr)$ of $\algs$ for $\ommt(U)$ \textsf{simulates} $H$ if $\cavity(H)=\cavity(H')$ and $\ring{d}_T(H)=\ring{d}_U(H')$.
\end{definition}

The following lemma is helpful for reducing the analysis of hybrid instances with biased cavity positions to the analysis of hybrid instances of SD on trees with fewer vertices.
%
%
%
%
%
\begin{lemma}
\label{lem-simulate}
Let $H\in\mf{H}_T(h,\stsvr)$ be a hybrid instance of $\algs$ for $\ommt(T)$.

\begin{enumerate}

\item If there is no cavity of $H$ in $T_i$ for some $i=0,\ldots,l$, then there exists a hybrid instance $H'\in\mf{H}_{T_{-i}}(h,\stsvr)$ for $\ommt(T_{-i})$ that simulates $H$.

\item If all cavities of $H$ are in $T_i$ for some $i=0,\ldots,l$, then there exists a hybrid instance $H'\in\mf{H}_{T_{i}}(h,\stsvr)$ for $\ommt(T_{i})$ that simulates $H$.

\item If all cavities of $H$ are in $T^{(j)}$ for some $j=1,2$ and
there is no cavity of $H$ in $T_0$, then there exists a hybrid instance $H'\in\mf{H}_{T^{(j)}}(h,\stsvr)$ for $\ommt(T^{(j)})$ that simulates $H$.

\end{enumerate}
\end{lemma}

\begin{proof}
Since we can show (1), (2) and (3) analogously, we show only (1).
Let $H=(\hyb,I)=((\algs,\start,\stsvr),(T,\sigma))$, $\sigma=r_1\ldots r_k$ and $\tend$ be the {\etime} of $H$.
Without loss of generality,
suppose by Lemma~\ref{lem-well-behaved} that $H$ is well-behaved.
By the definition of well-behaved hybrid instances,
all servers in $V(T)\setminus \cavity(H)$ are filled with exactly one request in $\{r_1,\ldots,r_{\start-1}\}$ with zero cost.
Therefore, without loss of generality,
we can suppose that requests from $r_1$ to $r_{n_i}$ are assigned to servers in $T_i$,
where $n_i\coloneqq|V(T_i)|$.

We now construct $H'=(\hyb',I')$ for $\ommt(T_{-i})$, where $\hyb'=(\algs,\start',\stsvr)$ and $I'=(T_{-i},\sigma')$.
Let $\start'\coloneqq\start-n_i$ and
define $\sigma'=r'_1\ldots r'_{k-n_i}$ as follows: if $r_{n_i+t}$ occurs in $T_i$, then let $r'_t\coloneqq\parent(\rho_i)$, otherwise $r'_t\coloneqq r_{n_i+t}$. 

Clearly,
$F_{\start-1}(\algs,I)=F_{\start'-1}(\algs,I')$
and
$F_{\start-1}(\hyb,I)=F_{\start'-1}(\hyb',I')$ hold.
The definition of $\algs$ shows that
$\algs$ for $T$ is completely the same as $\algs$ for $T_{-i}$ when no free server is in $T_i$.
This implies that the assignment of $r_{\start},\ldots,r_k$ by $\algs$ (resp. $\hyb$)
is completely the same as that of
$r'_{\start'},\ldots,r'_{k-n_i}$ by $\algs$ (resp. $\hyb'$).
In other words, it follows that
$h_{\start+t}^H=h_{\start'+t}^{H'}$
and
$a_{\start+t}^H=a_{\start'+t}^{H'}$
for $0\leq t\leq \tend-\start$.
Hence, we have
$H'\in\mf{H}_{T_{-i}}(h,\stsvr)$, $\cavity(H)=\cavity(H')$ and $\ring{d}_T(H)=\ring{d}_{T_{-i}}(H')$.
\end{proof}

The following lemma claims that the range of cavities is restricted when the positions of the first $\alg$-cavity and the first $\hyb$-cavity satisfy a certain condition.
%
%
%
%
%
%
\begin{lemma}
\label{lem-ha-ti}
Let $H\in \mf{H}_T(h,\stsvr)$ be a hybrid instance of $\algs$ for $\ommt(T)$.
\begin{enumerate}

\item If $h,\stsvr\in T_i$ for some $i=0,\ldots,l$, then all cavities of $H$ are in $T_i$.

\item If $h,\stsvr\in T^{(j)}$ for some $j=1,2$ and there is no cavity of $H$ in $T_0$, then
all cavities of $H$ are in $T^{(j)}$.

\end{enumerate}
\end{lemma}

\begin{proof}
Since (1) and (2) are proved analogously, we prove only (1).
Let $H=(\hyb,I)=((\algs,\start,\stsvr),(T,\sigma))$ and $\sigma=r_1\ldots r_k$.
By contradiction, assume that $h_{t-1},a_{t-1}\in T_i$ and $h_t\notin T_i$ for some $\start+1\leq t \leq \tend$. In this case, by (1) of Proposition~\ref{prop-atht}, $a_t=a_{t-1}$ holds. 

If $r_t \in T_i$, then $\algs$ assigns $r_t$ to a server in $T_i$ according to $\algs_i$ since $a_t\in T_i$ is free.
However, by (2) of Proposition~\ref{prop-atht}, $\algs$ assigns $r_t$ to $h_t\notin T_i$ and this is a contradiction.
If $r_t\notin T_i$, then $\algs$ prefers free servers in $T_i$ most among $\{T_j\}_{j=0}^l$ that contains a free server in it since $\hyb$ assigns $r_t$ to $h_{t-1}\in T_i$ (by (2) of Proposition~\ref{prop-atht}).
On the other hand, since $\algs$ assigns $r_t$ to $h_{t}\notin T_i$, $\algs$ does not prefer $T_i$ among $\{T_j\}_{j=0}^l$ that contains a free server in it and this is a contradiction.
Therefore, there is no $t$ such that $h_{t-1},a_{t-1}\in T_i$ and $h_t\notin T_i$.
If we assume that $h_{t-1},a_{t-1}\in T_i$ and $a_t\notin T_i$ for some $\start+1\leq t \leq \tend$, then we can also derive a contradiction in the same way as above.
Therefore, if $h,\stsvr\in T_i$, then all cavities of $H$ are in $T_i$.
\end{proof}

%
%
%
%
%
%

%
%

%
%
%
%
%

\section{Competitive analysis of Subtree-Decomposition}
\label{sec-analysis}

In this section, we derive the upper bound on the competitive ratio of the SD algorithm $\algs$.
To this end, the following lemma is important.

\begin{lemma}
\label{lem-main1}
Let $H=(\hyb,I)=((\algs,\start,\stsvr),(T,\sigma))$ be any hybrid instance for $\ommt(T)$ and
let $T_H$ be the minimal subtree of $T$ that contains all cavities of $H$. Then, we have
\begin{equation}
\label{ineq-lem-main1}
    \ring{d}_T(H)\leq 2\wE(T_H)-2 d_T(\lca(h_{\start},\stsvr),\rho_H),
\end{equation}
where $\rho_H=\rho(T_H)$ and $\wE(T_H)\coloneqq|E(T_H)|w_{T_H}^{\max}$.
\end{lemma}

The proof of the lemma is lengthy, making extensive use of the lemmas about hybrid algorithms that we have proved in previous sections. Therefore, for the sake of readability, we postpone the proof to Appendix~\ref{app-proof-main-lemma}.

In what follows, we prove two propositions and show that $\algs$ is T-strongly $3|E(T)|$-competitive for $\ommt(T)$.
\begin{proposition}
\label{prop-drs}
Consider $\algs$ for $\ommt(T)$. Let $s, s'$ be any vertex of $T$ such that $s\neq s'$.
For any $r\in V(T)$,
if $\algs$'s priority of $s$ for $r$ is higher than that of $s'$, then 
\[
d_T(r,s')\geq d_T^{\max}(r,s')\geq d_T^{\max}(s,s').
\]
\end{proposition}

\begin{proof}
The first inequality is obvious. Then, we prove the inequality $d_T^{\max}(r,s')\geq d_T^{\max}(s,s')$.
The proof is by induction on $|V(T)|$.
For the base case, i.e., $|V(T)|=2$, the proposition obviously holds.
For the inductive step, consider a power-of-two weighted tree $T$ with $k$ vertices and 
assume that the proposition holds for any power-of-two weighted tree with less than $k$ vertices.

First, suppose that $s'\in T_i$ and $s\notin T_i$ for some $i=1,\ldots,l$.
Then, by the definition of $\algs$, $r$ is not in $T_i$ and $d_T^{\max}(r,s')= w_T^{\max} = d_T^{\max}(s,s')$ holds.
Next, suppose that $s, s'\in T_i$ for some $i=1,\ldots,l$.
If $r\notin T_i$,
then
$d_T^{\max}(r,s')=w_T^{\max} \geq d_T^{\max}(s,s')$
holds.
If $r\in T_i$,
then
$d_{T_i}^{\max}(r,s')\geq d_{T_i}^{\max}(s,s')$
holds
by the induction hypothesis for $T_i$.
Therefore, we have
\[
d_T^{\max}(r,s')=d_{T_i}^{\max}(r,s')\geq d_{T_i}^{\max}(s,s')=d_{T}^{\max}(s,s').
\]
This completes the proof.
\end{proof}

\begin{proposition}
\label{prop-wtmaxha}
Let $H=((\algs,\start,\stsvr),(T,\sigma))$ be a hybrid instance for $\ommt(T)$. Then,
\[
d_T^{\max}(h_{\start},\stsvr)= w_{T_H}^{\max}.
\]
\end{proposition}
\begin{proof}
Let $h\coloneqq h_{\start}$.
We prove this by induction on $|V(T)|$.
For the base case, i.e., $|V(T)|=2$, the proposition obviously holds.
For the inductive step, consider a power-of-two weighted tree $T$ with $k$ vertices and 
assume that the proposition holds for any power-of-two weighted tree with less than $k$ vertices.

If $\stsvr\in T_i$ and $h\notin T_i$ for some $i\in 1,\ldots,l$, then $d_T^{\max}(h,\stsvr)= w_{T_H}^{\max}=w_T^{\max}$.
If $h,\stsvr\in T_i$ for some $i$,
by (1) of Lemma~\ref{lem-ha-ti}, all cavities of $H$ is in $T_i$.
Hence, by (2) of Lemma~\ref{lem-simulate}, there exists a hybrid instance $H_i$ for $\ommt(T_i)$ that simulates $H$.
Therefore, we have
\[
d_T^{\max}(h,\stsvr)= d_{T_i}^{\max}(h,\stsvr)=w_{T_{H_i}}^{\max}=w_{T_H}^{\max},
\]
where the second equality is due to the induction hypothesis for $\ommt(T_i)$ and
the last equality is due to $\cavity(H)=\cavity(H_i)$.
\end{proof}
%
%
%
%
%
%
\begin{theorem}
\label{thm-sd-t-strongly-3k-3-comp}
Subtree-Decomposition is T-strongly $(3k-3)$-competitive for $\ommt$ with $k$ servers.
\end{theorem}
%
%
%
%
%
\begin{proof}
Let $H=((\algs,\start,\stsvr),(T,\sigma))$ be any hybrid instance for $\ommt(T)$.
By Lemma~\ref{lem-main1} and Proposition~\ref{prop-wtmaxha}, we have
\[
    \ring{d}_T(H)
    \leq 2\wE(T_H)-2 d_T(\lca(h_{\start},\stsvr),\rho_H)
    \leq 2|E(T_H)|w_{T_H}^{\max}
    = 2|E(T_H)|d_{T}^{\max}(h_{\start},\stsvr).
\]
Since $\algs$ assigns $r_{\start}$ to not $\stsvr$ but $h_{\start}$, the $\algs$'s priority of $h_{\start}$ for $r_{\start}$ is higher than that of $\stsvr$.
Therefore, by Proposition~\ref{prop-drs}, we have
$d_{T}^{\max}(h_{\start},\stsvr)
\leq d_T^{\max}(r_{\start},\stsvr)$ and then
\[
    \ring{d}_T(H)
    \leq 2|E(T_H)|d_T^{\max}(r_{\start},\stsvr)
    \leq 2|E(T)|d_T^{\max}(r_{\start},\stsvr).
\]
By (2) of Corollary~\ref{cor-hyb-comp}, it follows that $\algs$ is T-strongly $3|E(T)|$-competitive for $\ommt(T)$.
By substituting $|E(T)|=k-1$, we can see that $\algs$ is T-strongly $(3k-3)$-competitive for $\ommt$ with $k$ servers.
\end{proof}
%
%
%
%
%
By applying Theorem~\ref{thm-suffice-for-omt-mpfs}, we obtain an $O(m)$-competitive algorithm for $\ofa(k,m)$.
\begin{corollary}
\label{cor-sd-ofa}
For $\ofa(k,m)$, there exists a deterministic $(12m-11)$-competitive algorithm.
\end{corollary}

In fact, with a more careful analysis, we can demonstrate that there exists a $(4k-3)$-competitive algorithm for $\omm(k)$ and an $(8m-5)$-competitive algorithm for $\ofa(k,m)$.

%
%
%
%
%
\begin{theorem}
\label{thm-sd-oms}
There exists a deterministic $(4k-3)$-competitive algorithm for $\omm(k)$.
\end{theorem}

\begin{proof}
Fix any instance $I=(S,d,\sigma)$ of $\omm(k)$ and
let $\mc{B}^*$ be the algorithm obtained by converting SD, as described in the proof of Theorem~\ref{thm-storongly-competitive-to-genera-lmetrics}.
Recall that $\mc{B}^*$ first constructs a power-of-two weighted tree $T$ by using the given metric space $(S,d)$ and then simulates SD with an instance $I'=(T,\sigma)$.
Let $H=((\mc{B}^*,\start,\stsvr),(S,d,\sigma))$ be a hybrid instance of $\mc{B}^*$ for $\omm(k)$ and $H'=((\algs,\start,\stsvr),(T,\sigma))$ be a hybrid instance of $\algs$ for $\ommt(T)$.
Then, we have
\begin{align*}
\ring{d}(H)&\leq \ring{d}_T(H')\leq 2\wE(T_{H'})= 2|E(T_{H'})|w_{T_{H'}}^{\max} \\
&\leq 2|E(T)|d_T^{\max}(r_{\start},\stsvr) \leq 4|E(T)|d(r_{\start},\stsvr)= (4k-4)d(r_{\start},\stsvr),
\end{align*}
where the first and fourth inequality is due to Claim~\ref{claim-sd-dist},
the second inequality is due to Lemma~\ref{lem-main1} and
the third inequality is due to Proposition~\ref{prop-wtmaxha} and $E(T_{H'})\subseteq E(T)$.
By (1) of Corollary~\ref{cor-hyb-comp}, this implies that $\mc{B}^*$ is $(4k-3)$-competitive.
\end{proof}
%
%
%
%
%
By applying Theorems~\ref{thm-server-eq} and~\ref{thm-mpfs}, we obtain an $(8m-5)$-competitive algorithm for $\ofa(k,m)$.

\begin{theorem}
\label{thm-sd-8m-5}
There exists a deterministic $(8m-5)$-competitive algorithm for $\ofa(k,m)$.
\end{theorem}

%
%

%
%
%
%
%

\section{Conclusion}
\label{sec-conclusion}

In this paper, we addressed the online transportation problem $\ofa(k,m)$ with $k$ servers located at $m$ server sites.

We first defined the T-strong competitive ratio for a tree metric and demonstrate a generic method for converting an algorithm designed for tree metrics into one for general metric spaces.
We believe that this technique can be applied to the open problem of designing an $O(\log k)$-competitive randomized algorithm for $\ommn(k)$ and so on.
 
We then introduced a new algorithm called Subtree-Decomposition and proved that
it is T-strongly $(3k-3)$-competitive
for $\ommt(T)$ with $k$ servers (in Theorem~\ref{thm-sd-t-strongly-3k-3-comp}).
We also demonstrated the existence of $(4k-3)$-competitive algorithms
for $\omm(k)$ (in Theorem~\ref{thm-sd-oms})
and $(8m-5)$-competitive algorithms for $\ofa(k,m)$
(in Theorem~\ref{thm-sd-8m-5}).
For any deterministic algorithm of $\ofa(k,m)$,
this upper bound on the competitive ratio is tight up to a constant factor with respect to its dependence on the number of server sites.
We conjecture that there exists an MPFS algorithm for $\ofa(k,m)$
that achieves the matching upper bound of $2m-1$.

\bibliography{ref}

\appendix
\section{Proof of Lemma~\ref{lem-main1}}
\label{app-proof-main-lemma}

Let $h\coloneqq h_{\start}$.
The proof is by induction on $|V(T)|$.
If $|V(T)|=2$, then the lemma obviously holds.
For the inductive step, consider a power-of-two weighted tree $T$ with $k$ vertices and 
assume that the lemma holds for any power-of-two weighted tree with less than $k$ vertices.
Fix any well-behaved hybrid instance $H=((\algs, \start,\stsvr),(T, \sigma))$ for $\ommt(T)$.
We break the proof into the following four cases:

\begin{itemize}
\setlength{\leftskip}{1.0cm}
\item[Case 1:] $H$ has no cavity in $T_i$ for some $0\leq i \leq l$,
\item[Case 2:] $h\in T_0$ and $\stsvr\in T_i$ for some $1\leq i \leq l$,
\item[Case 3:] $h\in T_i$ and $\stsvr\in T_j$ for some $1\leq i,j \leq l$ $(i\neq j)$ and
\item[Case 4:] $h\in T_i$ and $\stsvr\in T_0$ for some $1\leq i \leq l$
\end{itemize}
Note that if $h,\stsvr\in T_i$ for some $i$, then all cavities are in $T_i$ (by Lemma~\ref{lem-ha-ti}). Thus, that case is included in Case 1.
To prove this lemma, we use the following propositions
whose proofs can be found in Appendix~\ref{app-pf-prop-tree-metric} and~\ref{app-pf-prop-main2} respectively.
\begin{proposition}
\label{prop-tree-metric}
Let $T=(V(T),E(T))$ be a rooted tree with a weight function $w_T:E(T)\to \mathbb{R}_{\geq 0}$. For any vertex $v_1,v_2,v_3,v_4\in T$ and any common ancestor $\rho$ of $v_1,v_2,v_3$ and $v_4$, we have
\begin{align}
-d_T(v_1&,v_2)-d_T(v_3,v_4)+d_T(v_1,v_3)+d_T(v_2,v_4)\nonumber \\
\label{eq-tree-metric}
&= 2d_T(\ell_{1,2},\rho)+2d_T(\ell_{3,4},\rho)-2d_T(\ell_{1,3},\rho)-2d_T(\ell_{2,4},\rho), 
\end{align}
where $d_T$ denotes the path distance on $T$ and $\ell_{i,j}\coloneqq\lca(v_i,v_j)$ for any $1\leq i<j\leq 4$.
\end{proposition}
%
%
%
%
%
%
%
\begin{proposition}
\label{prop-main2}
Let $H=(\hyb,I)=((\algs,\start,\stsvr),(T,\sigma))$ be any hybrid instance for $\ommt(T)$ and
let $T_H$ be the minimal subtree of $T$ which contains all cavities of $H$.
If $H$ has no cavity in $T_0$, then we have
\begin{equation}
\label{ineq-prop-main2}
\ring{d}_T(H)\leq 2\wE(T_H\setminus T_0)+2\wE(T_H\cap T_0)-2 d_T(\lca(h_{\start},\stsvr),\rho_H).
\end{equation}
\end{proposition}
%
%
%
%
\half
\noindent\textbf{Case 1: \boldmath{$H$} has no cavity in \boldmath{$T_i$} for some \boldmath{$0\leq i \leq l$}.}
\half

By the definition of $\wE(\cdot)$, we have
\[
\wE(T_H\setminus T_0)+\wE(T_H\cap T_0)\leq \wE(T_H).
\]
If $i=0$, then by applying Proposition~\ref{prop-main2} to $H$, we have
\begin{align*}
\ring{d}_T(H)&\leq 2\wE(T_H\setminus T_0)+2\wE(T_H\cap T_0)-2 d_T(\lca(h,\stsvr),\rho_H)\\
&\leq 2\wE(T_H)-2 d_T(\lca(h,\stsvr),\rho_H).
\end{align*}

\noindent
If $i\neq 0$, then there exists a hybrid instance $H'$ for $\ommt(T_{-i})$ that simulates $H$ (by Lemma~\ref{lem-simulate}). By the induction hypothesis for $T_{-i}$, we have
\begin{align*}
\ring{d}_T(H)&=\ring{d}_{T_{-i}}(H')\leq
2\wE(T_{H'})-2 d_T(\lca(h,\stsvr),\rho_{H'})\\
&= 2\wE(T_H)-2 d_T(\lca(h,\stsvr),\rho_H),
\end{align*}
where the last equality is due to $\cavity(H)=\cavity(H')$.

%
%
%
%
\half
\noindent\textbf{Case 2: \boldmath{$h\in T_0$} and \boldmath{$\stsvr\in T_i$} for some \boldmath{$1\leq i \leq l$}.}
\half

In this case, we further divide the proof into the following four cases (of which only Case 2-1 and Case 2-4 are essentially important):

\begin{itemize}
\setlength{\leftskip}{2.0cm}

\item[Case 2-1:] There exists a time $u$ such that $h_u\in T_0$ and $h_{u+1}\in T_i$,

\item[Case 2-2:] There exists a time $u$ such that $a_u\in T_i$ and $a_{u+1}\in T_0$,

\item[Case 2-3:] All $\hyb$-cavities (resp. $\algs$-cavities) are in $T_0$ (resp. $T_i$) and

\item[Case 2-4:] There exists a time $u$ such that $h_u\in T_0$ and $h_{u+1}\in T_j$ ($j\neq i, j\neq 0$).

\end{itemize}

\half
\noindent\textbf{Case 2-1: There exists a time \boldmath{$u$} such that \boldmath{$h_u\in T_0$} and \boldmath{$h_{u+1}\in T_i$}.}
\half

By Lemma~\ref{lem-hybrid-main1}, there exists a hybrid instance $H_0\in\mf{H}_T(h,h_u)$ and $H_i\in\mf{H}_T(h_{u+1},\stsvr)$ such that $\cavity(H_0)=\{h_{\start},\ldots, h_u\}$, $\cavity(H_i)=\{h_{u+1},\ldots,h_{\tend},a_{\tend},\ldots,a_{\start}\}$,
\[
\ring{d}_T(H_0)=\ring{d}_T(h_{\start},\ldots, h_u) \text{ and }
\ring{d}_T(H_i)=\ring{d}_T(h_{u+1},\ldots,h_{\tend},a_{\tend},\ldots,a_{\start}).
\]
In addition, by Lemma~\ref{lem-simulate}, there exists a hybrid instance $H'_0$ (resp. $H'_i$) for $\ommt(T_0)$ (resp. $\ommt(T_i)$) that simulates $H_0$ (resp. $H_i$).
For clarity, we use the following notation:
\begin{itemize}
\item $\ell_{x,y}\coloneqq\lca(x,y)$ for any $x,y\in V(T)$,
\item $\rho'_0\coloneqq\rho(T_{H'_0})(=\rho(T_{H_0}))$ and 
\item $\rho'_i\coloneqq\rho(T_{H'_i})(=\rho(T_{H_i}))$.
\end{itemize}
By the definition of $\algs$, it is easy to see that $\rho'_0=\rho_H$.
Then, we have
\begin{align*}
\ring{d}_T(H)&=\ring{d}_T(h_{\start},\ldots, h_u)+\ring{d}_T(h_{u+1},\ldots,a_{\start}) \\
&\;\;\;\;\;\; -d_T(h,h_u)-d_T(h_{u+1},\stsvr)+d_T(h_u,h_{u+1})+d_T(h,\stsvr) \\
&=\ring{d}_{T_0}(H'_0)+\ring{d}_{T_i}(H'_i) \\
&\;\;\;\;\;\; +2d_T(\ell_{h,h_u},\rho_H)+2d_T(\ell_{h_{u+1},\stsvr},\rho_H)-2d_T(\ell_{h_u,h_{u+1}},\rho_H)-2d_T(\ell_{h,\stsvr},\rho_H) \\
&\leq 2\wE(T_{H'_0})-2d_T(\ell_{h,h_u},\rho'_0)+2\wE(T_{H'_i})-2d_T(\ell_{h_{u+1},\stsvr},\rho'_i)\\
&\;\;\;\;\;\; +2d_T(\ell_{h,h_u},\rho_H)+2d_T(\ell_{h_{u+1},\stsvr},\rho_H)-2d_T(\ell_{h_u,h_{u+1}},\rho_H)-2d_T(\ell_{h,\stsvr},\rho_H) \\
&= 2\wE(T_{H'_0})+2\wE(T_{H'_i})+2d_T(\rho'_i,\rho_H)-2d_T(\ell_{h_u,h_{u+1}},\rho_H)-2d_T(\ell_{h,\stsvr},\rho_H) \\
&\leq 2\wE(T_{H'_0})+2\wE(T_{H'_i})+2d_T(\rho'_i,\rho_H)-2d_T(\ell_{h,\stsvr},\rho_H).
\end{align*}
where the second equality is due to Proposition~\ref{prop-tree-metric}, the first inequality is due to the induction hypothesis for $T_0$ and $T_i$ and the third equality is due to $\rho'_0=\rho_H$ and $d_T(\ell_{h_{u+1},\stsvr},\rho_H)-d_T(\ell_{h_{u+1},\stsvr},\rho'_i)=d_T(\rho'_i,\rho_H)$.
Let $v$ be the vertex of $T_{H'_0}$ closest to $\rho'_i$, i.e., $T_H$ consists of $T_{H'_0}$, $T_{H'_i}$ and $\rho'_i$-$v$ path.
This implies that
\[
|E(T_H)|=|E(T_{H'_0})|+|E(T_{H'_i})|+|P_T(\rho'_i,v)|.
\]
Since $T_0$ does not contain any edge in $E_{\max}(T)$ and $T_H$ contains at least one edge in $E_{\max}(T)$, we obtain
$2w_{T_0}^{\max}\leq w_{T_H}^{\max}$ (recall $T$ to be a power-of-two weighted tree). Therefore, we have
\begin{align*}
\wE(&T_{H'_0})+\wE(T_{H'_i})+d_T(\rho'_i,\rho_H) \\
&=w_{T_{H'_0}}^{\max}|E(T_{H'_0})|+w_{T_{H'_i}}^{\max}|E(T_{H'_i})|+d_T(\rho'_i,v)+d_T(v,\rho_H) \\
&\leq w_{T_{0}}^{\max}|E(T_{H'_0})|+w_{T}^{\max}|E(T_{H'_i})|+w_{T}^{\max}|E(P_T(\rho'_i,v))|+w_{T_0}^{\max}|E(P_T(v,\rho_H))| \\
&\leq 2w_{T_{0}}^{\max}|E(T_{H'_0})|+w_{T}^{\max}(|E(T_{H'_i})|+|E(P_T(\rho'_i,v))|) \\
&\leq w_{T}^{\max}(|E(T_{H'_0})|+|E(T_{H'_i})|+|E(P_T(\rho'_i,v))|) \\
&= w_T^{\max}|E(T_H)|=\wE(T_H).
\end{align*}
Putting the above all together, we finally obtain
\[
\ring{d}_T(H)\leq 2\wE(T_H)-2d_T(\lca(h,\stsvr),\rho_H).
\]

\half
\noindent\textbf{Case 2-2: There exists a time \boldmath{$u$} such that \boldmath{$a_u\in T_i$} and \boldmath{$a_{u+1}\in T_0$}.}
\half

By replacing $h_u$ with $a_{u+1}$ and $h_{u+1}$ with $a_u$ in the proof of Case 2-1, we can similarly show (\ref{ineq-lem-main1}).

\half
\noindent\textbf{Case 2-3: All \boldmath{$\hyb$}-cavities (resp. \boldmath{$\algs$}-cavities) are in \boldmath{$T_0$} (resp. \boldmath{$T_i$}).}
\half

By replacing $h_u$ with $h_{\tend}$ and $h_{u+1}$ with $a_{\tend}$ in the proof of Case 2-1, we can show (\ref{ineq-lem-main1}).

%
%
%
%
%
%
%
\half
\noindent\textbf{Case 2-4: There exists a time \boldmath{$u$} such that \boldmath{$h_u\in T_0$} and \boldmath{$h_{u+1}\in T_j$} (\boldmath{$j\neq i, j\neq 0$}).}
\half

By Lemma~\ref{lem-hybrid-main1}, there exists a hybrid instance $H_0\in\mf{H}_T(h,h_u)$ and $H_1\in\mf{H}_T(h_{u+1},\stsvr)$ such that $\cavity(H_0)=\{h_{\start},\ldots, h_u\}$, $\cavity(H_1)=\{h_{u+1},\ldots,h_{\tend},a_{\tend},\ldots,a_{\start}\}$,
\[
\ring{d}_T(H_0)=\ring{d}_T(h_{\start},\ldots, h_u) \text{ and }
\ring{d}_T(H_1)=\ring{d}_T(h_{u+1},\ldots,h_{\tend},a_{\tend},\ldots,a_{\start}).
\]
In addition, by Lemma~\ref{lem-simulate}, there exists a hybrid instance $H'_0$ for $\ommt(T_0)$ that simulates $H_0$. By the induction hypothesis for $\ommt(T_0)$, we have
\begin{equation}
\label{eq-a}
\ring{d}_{T_0}(H'_0)\leq 2\wE(T_{H_0})-2d_T(\lca(h,h_u),\rho_{H_0}).
\end{equation}
Note that there is no cavity of $H_1$ in $T_0$.
By applying Proposition~\ref{prop-main2} to $H_1$,
we obtain
\begin{equation}
\label{eq-b}
\ring{d}_T(H_1)\leq 2\wE(T_{H_1}\setminus T_0)+2\wE(T_{H_1}\cap T_0)-2 d_T(\lca(h_{u+1},\stsvr),\rho_{H_1}).
\end{equation}
By Proposition~\ref{prop-tree-metric}, (\ref{eq-a}) and (\ref{eq-b}), it follows that
\begin{align*}
\ring{d}_T(H)&=\ring{d}_T(h_{\start},\ldots, h_u)+\ring{d}_T(h_{u+1},\ldots,a_{\start}) \\
&\;\;\;\;\;\; -d_T(h,h_u)-d_T(h_{u+1},\stsvr)+d_T(h_u,h_{u+1})+d_T(h,\stsvr) \\
&=\ring{d}_{T_0}(H'_0)+\ring{d}_{T}(H_1) \\
&\;\;\;\;\;\; +2d_T(\ell_{h,h_u},\rho_H)+2d_T(\ell_{h_{u+1},\stsvr},\rho_H)-2d_T(\ell_{h_u,h_{u+1}},\rho_H)-2d_T(\ell_{h,\stsvr},\rho_H) \\
&\leq 2\wE(T_{H_0})-2d_T(\ell_{h,h_u},\rho_{H_0}) \\
&\;\;\;\;\;\;+2\wE(T_{H_1}\setminus T_0)+2\wE(T_{H_1}\cap T_0)-2 d_T(\lca(h_{u+1},\stsvr),\rho_{H_1})\\
&\;\;\;\;\;\; +2d_T(\ell_{h,h_u},\rho_H)+2d_T(\ell_{h_{u+1},\stsvr},\rho_H)-2d_T(\ell_{h_u,h_{u+1}},\rho_H)-2d_T(\ell_{h,\stsvr},\rho_H) \\
&\leq 2\wE(T_{H_0})+2\wE(T_{H}\setminus T_0)+2\wE(T_{H_1}\cap T_0) \\
&\;\;\;\;\;\;+2d_T(\rho_{H_0},\rho_H)+2d_T(\rho_{H_1},\rho_H)-2d_T(\ell_{h,\stsvr},\rho_H).
\end{align*}
In addition, we also have
\begin{align*}
\wE(T&_{H_0})+d_T(\rho_{H_0},\rho_H)+\wE(T_{H_1}\cap T_0)+d_T(\rho_{H_1},\rho_H)\\
&\leq w_{T_0}^{\max}(|E(T_{H_0})|+|P_T(\rho_{H_0},\rho_H)|)+w_{T_0}^{\max}(|E(T_{H_1}\cap T_0)|+|P_T(\rho_{H_1},\rho_H)|) \\
&\leq 2w_{T_0}^{\max}|E(T_H\cap T_0)|\leq w_T^{\max}|E(T_H\cap T_0)|,
\end{align*}
where the last inequality is due to the definition of power-of-two weighted trees.
By the above two inequalities, we finally obtain
\begin{align*}
\ring{d}_T(H)&\leq 2\wE(T_{H}\setminus T_0)+2w_T^{\max}|E(T_H\cap T_0)|-2d_T(\ell_{h,\stsvr},\rho_H)\\
&\leq 2\wE(T_H)-2d_T(\ell_{h,\stsvr},\rho_H).
\end{align*}

\half
\noindent\textbf{Case 3: \boldmath{$h\in T_i$} and \boldmath{$\stsvr\in T_j$} for some \boldmath{$1\leq i,j \leq l$ $(i\neq j)$}.}
\half

In this case, we need the following proposition on the assignment of $\algs$.
The proof is given in Appendix~\ref{app-pf-claim-lcars}.

\begin{proposition}
\label{prop-lcars}
Consider a time step when $\algs$ assigns a current request $r$ to a free server $s$, and assume that $\lca(r,s)$ is an ancestor of $r$. Then, for each server $s'$ such that $\lca(r,s')$ is a descendant of $\lca(r,s)$, $s'$ is full at this time. 
\end{proposition}

If there is no cavity of $H$ in $T_0$, then such case is included in Case 1.
Thus, we consider the following two cases:
\begin{itemize}
\setlength{\leftskip}{2.0cm}

\item[Case 3-1:] There exists a time $u$ such that $h_u\in T_i$ and $h_{u+1}\in T_0$,

\item[Case 3-2:] There exists a time $u$ such that $a_u\in T_j$ and $a_{u+1}\in T_0$,

\end{itemize}

\half
\noindent\textbf{Case 3-1: There exists a time \boldmath{$u$} such that \boldmath{$h_u\in T_i$} and \boldmath{$h_{u+1}\in T_0$}.}
\half

There exist hybrid instances $H_i$ for $\ommt(T_i)$ and $H_{-i}$ for $\ommt(T_{-i})$ such that
\begin{align*}
    \cavity(H_i)&=\{h_{\start},\ldots, h_u\}, \\
    \cavity(H_{-i})&=\{h_{u+1},\ldots,h_{\tend},a_{\tend},\ldots,a_{\start}\}, \\
    \ring{d}_T(H_i)&=\ring{d}_T(h_{\start},\ldots, h_u) \text{ and } \\
    \ring{d}_T(H_{-i})&=\ring{d}_T(h_{u+1},\ldots,h_{\tend},a_{\tend},\ldots,a_{\start})
\end{align*}
%
%
%
%
%
%
%
%
Then, we have
\begin{align*}
\ring{d}_T(H)&=\ring{d}_T(h_{\start},\ldots, h_u)+\ring{d}_T(h_{u+1},\ldots,a_{\start}) \\
&\;\;\;\;\;\; -d_T(h,h_u)-d_T(h_{u+1},\stsvr)+d_T(h_u,h_{u+1})+d_T(h,\stsvr) \\
&=\ring{d}_{T_i}(H_i)+\ring{d}_{T_{-i}}(H_{-i}) \\
&\;\;\;\;\;\; +2d_T(\ell_{h,h_u},\rho_H)+2d_T(\ell_{h_{u+1},\stsvr},\rho_H)-2d_T(\ell_{h_u,h_{u+1}},\rho_H)-2d_T(\ell_{h,\stsvr},\rho_H) \\
&\leq 2\wE(T_{H_i})-2d_T(\ell_{h,h_u},\rho_{H_i})+2\wE(T_{H_{-i}})-2 d_T(\ell_{h_{u+1},\stsvr},\rho_{H_{-i}})\\
&\;\;\;\;\;\; +2d_T(\ell_{h,h_u},\rho_H)+2d_T(\ell_{h_{u+1},\stsvr},\rho_H)-2d_T(\ell_{h_u,h_{u+1}},\rho_H)-2d_T(\ell_{h,\stsvr},\rho_H) \\
&= 2\wE(T_{H_i})+2\wE(T_{H_{-i}})+2d_T(\rho_{H_i},\ell_{h_u,h_{u+1}})-2d_T(\ell_{h,\stsvr},\rho_H),
\end{align*}
where the last equality is due to $\rho_{H_{-i}}=\rho_H$.

Consider the $(u+1)$-th request $r_{u+1}$ of $H=(\hyb,I)$.
By Proposition~\ref{prop-atht}, $\algs$ assigns $r_{u+1}$ to $h_{u+1}\in T_0$ and $\hyb$ assigns to $h_u\in T_i$. This implies that $r_{u+1}$ is in $T_i$.
Moreover, by Proposition~\ref{prop-lcars}, each server $s$ such that $\lca(r_{u+1},s)$ is a descendant of $\lca(r_{u+1}, h_{u+1})$ is full at time $u+1$. Therefore, we can see that the vertex of $T_{H_{-i}}$ closest to $\rho_{H_i}$ is $\lca(r_{u+1}, h_{u+1})=\lca(h_u,h_{u+1})=\ell_{h_u,h_{u+1}}$.
This means that $T_H$ consists of $T_{H_i}$, $T_{H_{-i}}$ and $\rho_{H_i}$-$\ell_{h_u,h_{u+1}}$ path, i.e., 
\[
|E(T_H)|=|E(T_{H_i})|+|E(T_{H_{-i}})|+|E(P_T(\rho_{H_i},\ell_{h_u,h_{u+1}}))|.
\]
Then, we finally obtain
\[
\wE(T_{H_i})+\wE(T_{H_{-i}})+d_T(\rho_{H_i},\ell_{h_u,h_{u+1}})\leq \wE(T_H)
\]
and
\[
\ring{d}_T(H)\leq 2\wE(T_H)-2d_T(\ell_{h,\stsvr},\rho_H).
\]

\half
\noindent\textbf{Case 3-2: There exists a time \boldmath{$u$} such that \boldmath{$a_u\in T_j$} and \boldmath{$a_{u+1}\in T_0$}.}
\half

Let $H'$ be a conjugate instance of $H$.
Then, we have
$h_{u}^{H'}=a_u\in T_j$, $h_{u+1}^{H'}=a_{u+1}\in T_0$,
$\ring{d}(H)=\ring{d}(H')$ and
$\cavity(H)=\cavity(H')$ (i.e., $T_H=T_{H'}$).
By applying the proof of Case 3-1 to $H'$, we obtain
\begin{align*}
\ring{d}_T(H)&=\ring{d}_T(H') \\
&\leq 2\wE(T_{H'})-2d_T(\lca(h_{\start}^{H'},a_{\start}^{H'}),\rho_{H'})\\
&=2\wE(T_H)-2d_T(\ell_{h,\stsvr},\rho_H).
\end{align*}

\half
\noindent\textbf{Case 4: \boldmath{$h\in T_i$} and \boldmath{$\stsvr\in T_0$} for some \boldmath{$1\leq i \leq l$}.}
\half

Let $H'$ be a conjugate instance of $H$.
Then, it follows that
$h_{\start}^{H'}=\stsvr \in T_0$, $a_{\start}^{H'}=h \in T_i$,
$\ring{d}(H)=\ring{d}(H')$ and
$\cavity(H)=\cavity(H')$ (i.e., $T_H=T_{H'}$).
By applying the proof of Case 2 to $H'$, we have
\begin{align*}
\ring{d}_T(H)&=\ring{d}_T(H') \\
&\leq 2\wE(T_{H'})-2d_T(\lca(h_{\start}^{H'},a_{\start}^{H'}),\rho_{H'})\\
&=2\wE(T_H)-2d_T(\ell_{h,\stsvr},\rho_H).
\end{align*}
This completes the proof of Lemma~\ref{lem-main1}.


\subsection{Proof of Proposition~\ref{prop-tree-metric}}
\label{app-pf-prop-tree-metric}

Let $d_{i,j}\coloneqq d_T(v_i,v_j)$ and $d_i\coloneqq d_T(v_i,\rho)$.
By the property of tree metrics, we have
\[
2d_T(\ell_{i,j}, \rho)=d_T(v_i,\rho)+d_T(v_j,\rho)-d_T(v_i,v_j)=d_i+d_j-d_{i,j}.
\]
By substituting the above equation into the right hand side of (\ref{eq-tree-metric}), we have
\begin{align*}
\text{RHS of (\ref{eq-tree-metric})} &= (d_1+d_2-d_{1,2})+(d_3+d_4-d_{3,4})-(d_1+d_3-d_{1,3})-(d_2+d_4-d_{2,4}) \\
&=-d_{1,2}-d_{3,4}+d_{1,3}+d_{2,4} = \text{LHS of (\ref{eq-tree-metric})}.
\end{align*}

\subsection{Proof of Proposition~\ref{prop-main2}}
\label{app-pf-prop-main2}

Let $h\coloneqq h_{\start}$.
The proof is by induction on $|V(T)|$.
if $|V(T)|=3$, then the proposition is trivial.
For the inductive step, consider a power-of-two weighted tree $T$ with $k$ vertices and 
assume the proposition holds for any power-of-two weighted tree with less than $k$ vertices.
For $j=1,2$ let $T_0^{(j)}\coloneqq(V(T^{(j)})\cap V(T_0), E(T^{(j)})\cap E(T_0))$.
Fix any well-behaved hybrid instance $H=((\algs, \start,\stsvr),(T, \sigma))$ for $\ommt(T)$ that has no cavity in $T_0$.
We consider the following five cases (of which only Case 1 and Case 2-1 are of essential importance):

\begin{itemize}
\setlength{\leftskip}{1.0cm}
\item[Case 1:] $h,\stsvr\in T^{(j)}$ for some $j=1,2$,

\item[Case 2:] $h\in T^{(1)}$ and $\stsvr\in T^{(2)}$,
\begin{itemize}
\setlength{\leftskip}{1.0cm}
\item[Case 2-1:] There exists a time $u$ such that $h_u\in T^{(1)}$ and $h_{u+1}\in T^{(2)}$,
\item[Case 2-2:] There exists a time $u$ such that $a_u\in T^{(2)}$ and $a_{u+1}\in T^{(1)}$,
\item[Case 2-3:] All $\hyb$-cavities (resp. $\algs$-cavities) are in $T^{(1)}$ (resp. $T^{(2)}$) and
\end{itemize}

\item[Case 3:] $h\in T^{(2)}$ and $\stsvr\in T^{(1)}$.

\end{itemize}

\noindent \textbf{Case 1: \boldmath{$h,\stsvr\in T^{(j)}$} for some \boldmath{$j=1,2$}.}
\half
In this case, all cavities of $H$ are in $T^{(j)}$ by Lemma~\ref{lem-ha-ti}.
Then, by Lemma~\ref{lem-simulate}, there exists a hybrid instance $H'$ for $\ommt(T^{(j)})$ that simulates $H$.
By the induction hypothesis for $T^{(j)}$, we have
\begin{align*}
\ring{d}_T(H)&=\ring{d}_{T^{(j)}}(H') \\
&\leq 2\wE(T_{H'}\setminus T_0^{(j)})+2\wE(T_{H'}\cap T_0^{(j)})-2d_T(\lca(h,\stsvr),\rho_{H'}) \\
&=2\wE(T_{H}\setminus T_0^{(j)})+2\wE(T_H\cap T_0^{(j)})-2d_T(\lca(h,\stsvr),\rho_H) \\
&= 2\wE(T_{H}\setminus T_0)+2\wE(T_H\cap T_0)-2d_T(\lca(h,\stsvr),\rho_H).
\end{align*}

%
%
%
%
%
\noindent \textbf{Case 2-1: There exists a time \boldmath{$u$} such that \boldmath{$h_u\in T^{(1)}$} and \boldmath{$h_{u+1}\in T^{(2)}$}.}
\half
Note that $\rho_H=\rho$ in this case.
By Lemma~\ref{lem-hybrid-main1}, there exist hybrid instances $H_1\in\mf{H}_T(h,h_u)$ and $H_2\in\mf{H}_T(h_{u+1},\stsvr)$ such that
\begin{align*}
    \cavity(H_1)&=\{h_{\start},\ldots, h_u\}, \\
    \cavity(H_2)&=\{h_{u+1},\ldots,h_{\tend},a_{\tend},\ldots,a_{\start}\}, \\
    \ring{d}_T(H_1)&=\ring{d}_T(h_{\start},\ldots, h_u) \text{ and } \\
    \ring{d}_T(H_2)&=\ring{d}_T(h_{u+1},\ldots,h_{\tend},a_{\tend},\ldots,a_{\start}).
\end{align*}
In addition, by Lemma~\ref{lem-simulate}, there exists a hybrid instance $H'_1$ (resp. $H'_2$) for $\ommt(T^{(1)})$ (resp. $\ommt(T^{(2)})$) that simulates $H_1$ (resp. $H_2$).
For clarity, we use the following notation:
\begin{itemize}
\item $\ell_{x,y}\coloneqq\lca(x,y)$ for $x,y\in V(T)$,
\item $\rho'_1\coloneqq\rho(T_{H'_1})=\rho(T_{H_1})$ and
\item $\rho'_2\coloneqq\rho(T_{H'_2})=\rho(T_{H_2})$.
\end{itemize}
By the definition of $\algs$ in Phase 2, $T_H$ consists of $T_{H'_1}$, $T_{H'_2}$ and the $\rho'_1$-$\rho'_2$ path $P_T(\rho'_1,\rho'_2)$.
Then, we have
\begin{align}
\label{rhoh}
&\rho_H=\rho(T_H)=\rho, \\
\label{weth1setminus}
&\wE(T_{H'_1}\setminus T_0^{(1)})+\wE(T_{H'_2}\setminus T_0^{(2)})\leq \wE(T_H\setminus T_0) \text{ and } \\
\label{weth1cap}
&\wE(T_{H'_1}\cap T_0^{(1)})+\wE(T_{H'_2}\cap T_0^{(2)})+d_T(\rho'_1,\rho)+d_T(\rho'_2,\rho)\leq \wE(T_H\cap T_0).
\end{align}
Therefore, we finally get
\begin{align*}
\ring{d}_T(H)&=\ring{d}_T(h_{\start},\ldots, h_u)+\ring{d}_T(h_{u+1},\ldots,a_{\start}) \\
&\;\;\;\;\;\; -d_T(h,h_u)-d_T(h_{u+1},\stsvr)+d_T(h_u,h_{u+1})+d_T(h,\stsvr) \\
&=\ring{d}_{T^{(1)}}(H'_1)+\ring{d}_{T^{(2)}}(H'_2) \\
&\;\;\;\;\;\; +2d_T(\ell_{h,h_u},\rho)+2d_T(\ell_{h_{u+1},\stsvr},\rho)-2d_T(\ell_{h_u,h_{u+1}},\rho)-2d_T(\ell_{h,\stsvr},\rho) \\
&\leq 2\wE(T_{H'_1}\setminus T_0^{(1)})+2\wE(T_{H'_1}\cap T_0^{(1)})-2d_T(\ell_{h,h_u},\rho'_1) \\
&\;\;\;\;\;\; +2\wE(T_{H'_2}\setminus T_0^{(2)})+2\wE(T_{H'_2}\cap T_0^{(2)})-2d_T(\ell_{h_{u+1},\stsvr},\rho'_2)\\
&\;\;\;\;\;\; +2d_T(\ell_{h,h_u},\rho)+2d_T(\ell_{h_{u+1},\stsvr},\rho)-2d_T(\ell_{h_u,h_{u+1}},\rho)-2d_T(\ell_{h,\stsvr},\rho) \\
&= 2\wE(T_{H'_1}\setminus T_0^{(1)})+2\wE(T_{H'_1}\cap T_0^{(1)})+2d_T(\rho'_1,\rho) \\
&\;\;\;\;\;\; +2\wE(T_{H'_2}\setminus T_0^{(2)})+2\wE(T_{H'_2}\cap T_0^{(2)})+2d_T(\rho'_2,\rho)\\
&\;\;\;\;\;\; -2d_T(\ell_{h_u,h_{u+1}},\rho)-2d_T(\ell_{h,\stsvr},\rho) \\
&\leq 2\wE(T_H\setminus T_0) +2\wE(T_H\cap T_0) -2d_T(\ell_{h,\stsvr},\rho_H),
\end{align*}
where the second equality is due to Proposition~\ref{prop-tree-metric}, the first inequality is due to the induction hypothesis for $T^{(1)}$ and $T^{(2)}$ and the last inequality is due to the formulae (\ref{rhoh})-(\ref{weth1cap}).

\half
\noindent\textbf{Case 2-2: There exists a time \boldmath{$u$} such that \boldmath{$a_u\in T^{(2)}$} and \boldmath{$a_{u+1}\in T^{(1)}$}.}
\half

By replacing $h_u$ with $a_{u+1}$ and $h_{u+1}$ with $a_u$ in the proof of Case 2-1, we can similarly show (\ref{ineq-prop-main2}).

\half
\noindent\textbf{Case 2-3: All $\hyb$-cavities (resp. $\algs$-cavities) are in \boldmath{$T^{(1)}$} (resp. \boldmath{$T^{(2)}$)}.}
\half

By replacing $h_u$ with $h_{\tend}$ and $h_{u+1}$ with $a_{\tend}$ in the proof of Case 2-1, we can show (\ref{ineq-prop-main2}).

\half
\noindent\textbf{Case 3: \boldmath{$h\in T^{(2)}$} and \boldmath{$\stsvr\in T^{(1)}$}.}
\half

Let $H'$ be a conjugate instance of $H$. By the definition of conjugate instances, 
we have $h_{\start}^{H'}=\stsvr\in T^{(1)}$, $a_{\start}^{H'}=h\in T^{(2)}$,
$\cavity(H)=\cavity(H')$ (i.e., $T_H=T_{H'}$) and
$\ring{d}_T(H)=\ring{d}_T(H')$.
This implies that we can apply the proof of Case 2 for $H'$.
Thus, it follows that
\begin{align*}
\ring{d}_T(H)&=\ring{d}_T(H') \\
&\leq 2\wE(T_{H'}\setminus T_0)+2\wE(T_0)-2 d_T(\lca(h_{\start}^{H'},a_{\start}^{H'}),\rho_{H'}) \\
&=2\wE(T_H\setminus T_0)+2\wE(T_0)-2 d_T(\lca(h,\stsvr),\rho_H).
\end{align*}
This completes the proof.

\subsection{Proof of Proposition~\ref{prop-lcars}}
\label{app-pf-claim-lcars}

The proof is by induction on $|V(T)|$.
For the base case, i.e., $|V(T)|=2$ with $r \neq s$, the only server suitable for $s'$ in the claim is the position of $r$.
Then, $s'$ is obviously full if $r$ is assigned to $s$.
For the inductive step, we break the proof into three cases: (1) $r,s\in T_i$ for some $i=0,\ldots,l$, (2) $r\in T_i$, $s\in T_j$ ($i\neq j$) and $r,s\in T^{(j)}$ for some $j=1,2$ and (3) $r\in T^{(j)}$ and $s\in T^{(3-j)}$ for some $j=1,2$.

For (1), $\algs$ assigns $r$ to $s$ according to $\algs_i$. Then, by the induction hypothesis for $T_i$, the claim holds.
For (2), if $j=0$, then $\algs$ assigns $r$ to $s$ according to $\algs_0$ and all servers in $T_i$ are full. Thus, by the induction hypothesis for $T_0$, the claim holds.
If $j\neq 0$, then the claim can be shown by the induction hypothesis for $T^{(j)}$.
For (3), by the definition of $\algs$, all servers in $T^{(j)}$ is full at this time and the claim holds.

Hence, it is shown that the claim holds in all cases (1), (2) and (3).

\end{document}